\documentclass[12pt]{article}

\title{Visualizing Treasury Issuance Strategy}
\author{Christopher Cameron 
\thanks{The author is with the U.S. Department of the Treasury, Office of Debt Management. The 
analysis and 
conclusions set forth in this paper are those of the author, and do not necessarily reflect those of or 
indicate concurrence by other members of the Treasury staff, Treasury's senior officials, the Treasury Department, or the United 
States government.}
\\ \textit{christopher.cameron@treasury.gov}
}
\date{February 20, 2018 (This version: January 27, 2020)}

\usepackage[english]{babel}
\usepackage{amsmath,amsthm,tabu}
\usepackage{amsfonts}
\usepackage{breqn}
\usepackage{graphicx}
\usepackage[ampersand]{easylist}
\usepackage{booktabs}
\usepackage{multirow}
\usepackage{caption}
\usepackage{mathtools}

\DeclarePairedDelimiter{\floor}{\lfloor}{\rfloor}
\theoremstyle{definition}

\theoremstyle{remark}
 
\numberwithin{equation}{section}

\def\myhed{\medskip\noindent\textbf}
 
\begin{document}

\maketitle

\section{Introduction}

It is a major challenge in US Treasury debt management to communicate the stance and implications of current and contemplated debt-issuance actionably yet concisely. A complex, operationally-intensive policy that currently involves over 250 regular auctions of 14 security types annually, in various sizes and patterns totaling (as of 2017) almost \$4 trillion in total borrowing (here defined as net new debt
plus refinancing of prior-year debt), must be apprehended by strategists and advisors when considering what shall be done and describing the expected result. Discussants must summarize their views and findings in a common language, and also in a way that aligns with and informs the practical, on-the-ground decision points that policy makers actually control. 

Similarly, while there is a consensus to ensure that decisions be informed by (often quite complex) quantitative models and simulations created by researchers, it is no small task to bridge the gap between the product of such models and the tangible needs of and constraints on debt managers. What is needed is to align theory and numerical metrics with how the path of debt 
issuance is envisioned, discussed, and navigated by practitioners.

The purpose of this paper is to motivate and describe strategy metrics that help meet this need. These metrics and the model behind
them are derived from simple and intuitive reasoning about the long-term, asymptotic implications of ongoing debt issuance in a steady-state environment.
Because of the numerous simplifying assumptions required to render such a calculation coherent and well-defined, the 
underlying model is best 
understood as a complement to, and not a substitute for, the near- and medium-term debt simulation and optimization efforts
more commonly found in the
literature. 

By mapping issuance strategies according to these metrics, which (abstractly) represent cost and risk respectively, one
can trace the evolution of Treasury issuance historically, identify
an analogue of the efficient frontier, and gauge the likely impact of anticipated changes to the prevailing
issuance pattern. These visualizations can also be done on a forward-looking basis, and are based on the flow of new-issuance rather
than the outstanding stock. The resulting 
strategy metrics,
and visualizations based on them, can therefore help to summarize and illuminate Treasury issuance strategy in ways that aid debt managers in their
decisions in a way that is aligned with how those decisions are contemplated.

\section{Motivation}

In this section we describe same motivation for the development and usage of these metrics in visualizing Treasury issuance strategy.

\subsection{Escaping WAM}

The communication 
challenge described above is best exemplified by the pervasiveness of a single, high-level summary metric whenever discussion of debt management turns quantitative. This is the weighted-average maturity (WAM) of the outstanding portfolio. Often, this metric is taken to
serve as a simultaneous proxy for both cost and risk: all else equal, a portfolio with a high WAM would be expected to have 
high debt-servicing cost, and low risk (however measured, whether in terms of rollover amounts, debt-servicing cost volatility, or
other similar metrics).

\begin{figure}[htbp]
\includegraphics[scale=.7]{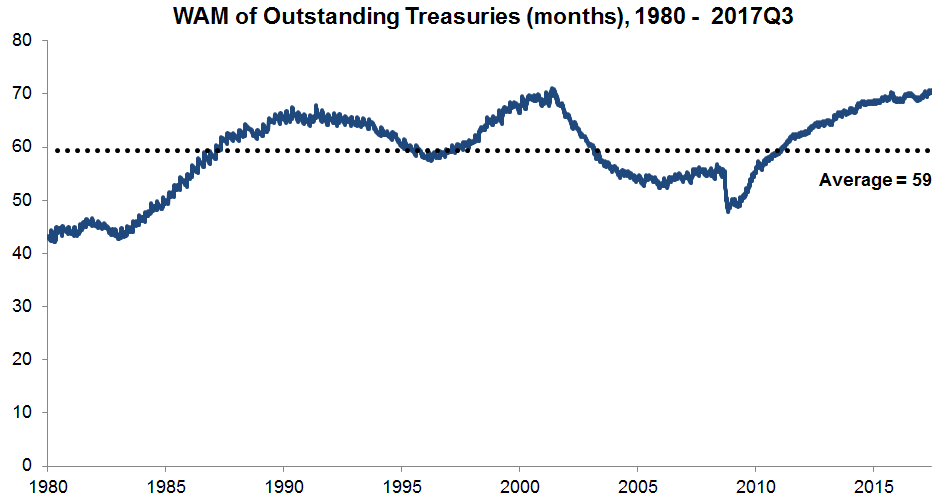}
\caption{WAM (weighted-average maturity) of Treasuries outstanding, including those held by the Federal Reserve. The current (2017Q3) WAM of roughly 71 months is near modern historical highs.  \label{fig_wam}  }
\end{figure}

\begin{figure}[htbp]
\includegraphics[scale=.7]{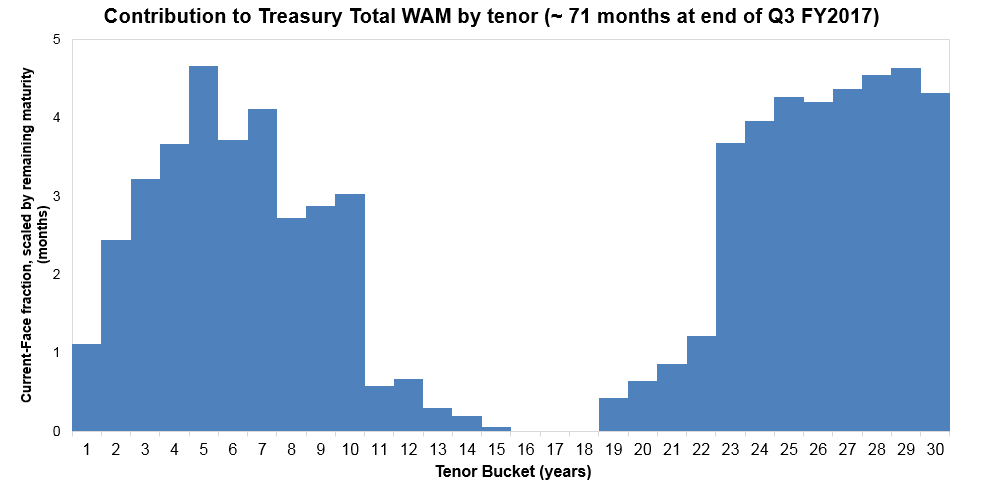}
\caption{An illustration of the distribution of outstanding Treasury securities, as regards their contribution to WAM. Here the current-face of each security is scaled by its remaining maturity (in months)
and divided by the total size of the portfolio;
the resulting quantities, having units of months, are then aggregated by maturity bucket. The sum of the column heights in the chart is the WAM itself. This visualization illustrates 
that the Treasury portfolio is essentially bifurcated into two distinct portfolios rather than uniformly allocated. The shorter ($\le 15$ years) portion of the portfolio has a WAM of around $33$ months, but the 
longer ($20+$ years) portion of the portfolio serves to boost WAM to its elevated level (approximately 71 months). All quantities as of end Q3 FY2017.  \label{fig_wamhistogram}  }
\end{figure}

It is true that the WAM of a portfolio can serve as useful and intuitive shorthand for both its cost and risk implications over time. The 
quantity is straightforward, transparently model-free for anyone to compute, and can be done so using public data. Yet all recognize that it has inherent shortcomings. For example, a "barbelled" portfolio composed of half $1$-day and half $30$-year bonds
has essentially the same WAM as a portfolio containing only $15$-year bonds, but very different rollover and risk implications. (In this
extreme example, half of the former portfolio will need to be refinanced in one day.) Similarly, the WAM of the US Treasury portfolio, 
which  is at or near modern-day highs (see Figure~\ref{fig_wam}), largely reflects increased issuance flow (in both
size and frequency) of longer-dated borrowing (7- and 10-year notes, and 30-year bonds) that began near the beginning of this decade. The resulting WAM of around 70 months is a consequence of disproportionately adding these longer-dated bonds to the portfolio (closely following the 2001-2006 discontinuation of 30-year issues), which can be seen to have a marked "hole" near maturities of around 15 to 20 years. (See Figure~\ref{fig_wamhistogram}.)

But WAM masks the fine-grained details of a maturity profile, and  to discuss debt management in terms of managing the portfolio WAM leads to oversimplification of what is in reality a dynamic, evolving process that is highly influenced by historical issuance patterns.

An added, less well appreciated observation is that WAM is a \emph{stock }measure (describing, after all, the \emph{outstanding} portfolio). Although such a proxy does fit with a 
rich literature in sovereign debt issuance modeling that addresses the optimal distribution of stock, 
in practice Treasury debt management choices and decisions (absent a significant buyback program) have a more direct
effect not on stock but on \emph{flow}. That is, under current policy
 it is (largely) the makeup and sizing of the regular, ongoing supply \emph{of new issuance} 
that is under the direct control of the 
US Treasury. This mismatch, between control variables and ostensible target metrics, can 
make certain statements and concepts phrased in terms of WAM misleading. 

\begin{figure}[htbp]
\includegraphics[scale=.55]{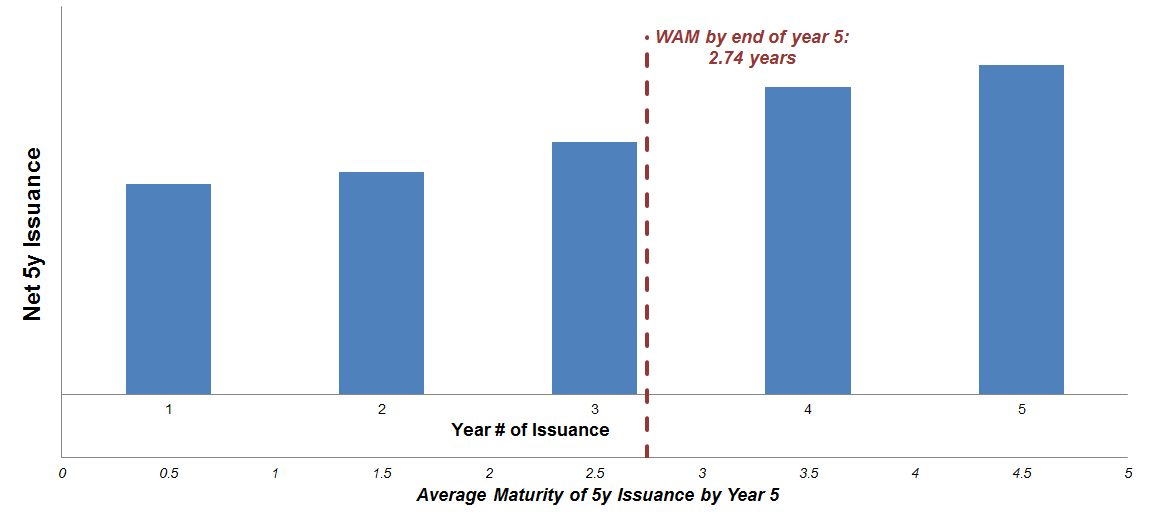}
\caption{Example of a hypothetical portfolio created by issuing only 5y debt yearly, spread throughout each year, in the face of
rising and randomly-fluctuating deficits. The average
maturity of a series of 5y notes \emph{with the same size} spread over five years  would be 2.5 years, but  because deficits grow
and fluctuate, the actual WAM of such a portfolio would fluctuate too, and likely be in the 2.5-3 year range.   \label{example5y}  }
\end{figure}

For example, consider a hypothetical strategy of using only 5-year issuance: this might sometimes be interpreted and spoken of (loosely) as 
using a "5 year WAM" strategy. But regular, periodic issuance of 5-year debt does \emph{not} create a portfolio with a 5-year WAM (the actual WAM of the resulting portfolio would likely range between 2.5-3 years, \emph{and fluctuate with deficits}). See 
Figure~\ref{example5y} for an illustration, which shows why this occurs and why there is a meaningful difference between the
maturity of new-issuance and the WAM of the resulting portfolio. It also illustrates that considering and managing the maturity \emph{of new debt} does not necessarily result in a predictable WAM, which can and will be affected by fluctuations in deficits and rates. (Conversely,
managing to a particular WAM would \emph{require allowing new issuance to fluctuate}  with deficits and rates.)

As of 2017, the Treasury portfolio has a WAM of almost 6 years (far longer than 3), issues at an average maturity of around 3.5-4 years (shorter than 5) and with 
a median maturity that is even shorter, at around 3 years. How then to translate a statement about something as simple as all-5-year issuance into the language of WAM? How to interpret its implications, or even speak about new-issuance allocation at all, when it is WAM that is the \textit{lingua franca} of debt management? Managing debt flow but measuring and tracking WAM becomes a regular source of miscommunication.

The truth is that despite these drawbacks, and however aware debt managers are of them and attempt to move beyond undue focus on WAM, its simplicity and ubiquity -- and, importantly, the lack of alternatives -- makes it “too useful” not to revert to as a strategy metric and shorthand, a common reference point for communicating strategy.  

\subsection{Flow and the cost-risk tradeoff}

As stated above, this piece describes an effort undertaken to improve this situation by summarizing and visualizing Treasury issuance in a way that, while it preserves much of the simplicity and transparency advantages of WAM, is more aligned with the dynamic and flow-centered nature of practical debt management and its tradeoffs. We do this by associating to any given issuance pattern or set of sizes two proxy metrics, one abstractly corresponding to cost and the other to risk. An issuance strategy or pattern is thereby represented as a point in a cost-versus-risk space (i.e., risk-return but from the issuer perspective). As patterns change or issue sizes are adjusted, the implied cost-risk tradeoff changes.  By plotting the trajectory over time, one can illustrate the path of Treasury debt management either historically, or in a given projected scenario. The resulting graph in cost-versus-risk space allows for visualization of debt-management strategy directions and movements in a way that harnesses intuitions and stylized facts from classical portfolio theory and efficient frontiers.

Often, scenario analysis in Treasury debt management involves charting out and considering \emph{a future path of issue-sizes} over a planning
period (for example, $10$ years). With over $250$ auctions per year, in the most general of such constructs there would be 
at least $2500$ 
independent control variables to consider. Simplification is inevitable. But even if planned representative issue-sizes in a forward 
planning period are (for example) stipulated to be  
constant yearly, with $14$ security types this still creates $140$ independent variables. Further reduction of the dimension of 
this problem is necessary to comprehensible scenario analysis or consideration of the impacts of a particular issuance strategy shift.

Practically speaking, then, scenario analysis largely confines itself to consideration of high-level, easily-comprehended
 strategy directions. To illustrate, suppose projected funding requirements compel Treasury to increase borrowing patterns versus
status quo. Various groupings of instruments, and/or tenor categories, may be considered, such as: increasing bill sizes pro rata,
increase coupon-bond sizes pro rata, twist short (so that short-tenor instrument sizes are increased faster than for long tenors), 
twist long, "barbell" (e.g. increase the very short and very long tenors, against the medium tenors), and so forth. The  time dimension involved in forward planning only complicates
matters, because (unlike with, say, equity allocation) the question is not how to allocate "right now" but how to do so \emph{over time}. And allocation is dynamic; it can and does change over time.

\subsection{Why focus on steady-state?}

It may be objected that a steady-state debt distribution equilibrium is unconvincing or unrealistic as a model of debt dynamics. It might fairly be expected that shocks, feedbacks, endogenous prices, and/or other more complex dynamics than are captured in this framework are what largely drive this or that salient conclusion to be drawn from more thorough and complex approaches to modeling debt issuance (we are agnostic, but admit to being unconvinced in that regard). While such critics are correct that
this framework is (unabashedly) a simplification, we hasten to add that our claims for the usefulness of such an approach are modest, and not tightly bound to any particular strong assertion about whether steady-state has been or will be, in fact, \emph{attained}. As stated above, an immediate goal is simply to develop a way of understanding and communicating intuitions about debt-dynamics, as they arise in planning, simulation, or modeling of debt issuance. To the extent those dynamics are driven at least in part -- perhaps in large part -- by the underlying mechanics of the evolution equations investigated below, surely one can only benefit from actually exploring and understanding the consequences of those equations, simplification though they are, to the full extent possible.

In addition, we observe that there appear to be certain characteristics shared by a large class of debt issuance models that readily and immediately motivate investigation in this direction. Consider models such as those developed in~\cite{bolder1} or~\cite{tbac}. What is, in fact, going in on such models? A rolling debt portfolio, under conditions of a well-defined prescribed issuance strategy, is simulated periodwise up to a large time horizon (say 20 or 30 years). The simulations are driven by an ensemble of input-paths (of interest rates, output gaps, deficits, and so on) drawn, implicitly, from a stochastic distribution calibrated against a history of economic conditions and constructed or constrained so as to be, it is believed, economically feasible. Metrics representing e.g. cost and risk are then observed in each simulation at or approaching the simulation horizon, and averaged over the ensemble of realizations. Finally, conclusions are drawn regarding the relationship between these long-term metrics (as estimated by the numerical simulation and averaging process) and the assumed issuance strategy, to derive conclusions or relationships regarding efficient or optimal policy directions. 

Mathematically, a model of this type can be thought of as  a \emph{Monte Carlo simulation}: a numerical estimation of the (perhaps incalculable analytically, but theoretically exact) mapping from the space of issuance strategies to the chosen metrics (i.e. distributional means or other moments of portfolio observables under the stochastic distribution of the input paths) at the long-time horizon chosen. This paper considers nothing other than the consequences of such a mapping as that long-time horizon, which after all is arbitrary, tends to infinity. Moreover, in most such models we have encountered, as can often be seen from their fan charts, the stochastic input paths driving simulations are quickly mean-reverting (e.g. Gaussian) to some apparently steady average, and not especially skewed, distribution, with little apparent influence from initial conditions. Note that the basic yearly budgeting equation governing debt dynamics, as usually expressed, is linear. All of this only buttresses the notion that the theoretical long-term steady-state of such a simulation is, to a significant degree, what is actually being approximated in such a simulation, and that the average metrics thus compiled are approximations to or at least well-proxied by the exact analytical metrics arising in a steady-state. (Indeed, were the closed-form steady-state formulas developed below unknown, a perfectly legitimate way to approximate them would be to run an ensemble of debt-simulations under steady or an ensemble of quickly-mean-reverting input paths and measure mean in-simulation observables at some large time-horizon.)

What follows then can be conceived of as an alternative approach to modeling or scenario analysis of the types referenced above, and one
that attempts to capture the salient features and expected long-term dynamics of more complex quantitative models (e.g. Monte Carlo methods that, in principle, estimate or at least approach a steady-state portfolio given market and strategy assumptions) that are, unavoidably, several orders of magnitude more computationally intensive. It is also a goal to obviate the need to focus on specific dollar 
issue sizes and near-term fluctuations along finely-calibrated forward market paths, and instead bring relative allocation, borrowing fractions, and underlying dynamics to the forefront
of strategic consideration.

Alternatively, and less ambitiously, this piece merely presents simple cost and risk proxy metrics relevant to and useful in visualizing issuance strategy, in the context of and as an adjunct to 
what is, in the abstract, a complex optimal stochastic control problem (for examples of the latter we can refer the reader to e.g.
\cite{bolder1}, \cite{consiglio}, \cite{date}). Note these metrics and formulas are substantively identical to those 
presented and discussed, with a 
slightly different emphasis, for the zero-debt-growth case in~\cite{landoni}. Equivalent formulas have also been more recently 
used to aid discussion
of modeling results by the Treasury Borrowing Advisory Committee in their presentations to Treasury, see for example PDF p. 68 of
~\cite{tbac_2018q4} or p. 112-113 of ~\cite{tbac_2019q2}. 

The next section gives definitions necessary to understanding of the proxy metrics we put forth.

\section{Definitions}
 
This section describes definitions and mathematical constructs required for the metrics we introduce. The
starting point is the government's periodwise budget 
equation, which (aside from, possibly, notation) is standard and does not depart from e.g. \cite{barro}.

Time is discretized into periods, which we will take to be (fiscal) years. In each year $t = 0, 1, 2, \dots$, dollar outlays are made from the Treasury General Account (TGA) broadly owing to three sources:
\begin{itemize}
\item The difference (a deficit, $D_t$) between fiscal spending and tax receipts. (This may or may not coincide with "primary deficit", as it 
includes \emph{all} (non-interest) spending items that may necessitate debt financing including e.g. student loans.)
\item Interest payments on debt, $I_t$
\item Repayment of maturing debt principal, $M_t$
\end{itemize}

To ensure that these outlays do not add to the money stock (a phenomenon that would register as a reduction in the 
TGA balance) they  must be offset by dollar inflows. This is achieved by issuing new debt in an amount $N_t$ sufficient to generate the required cash proceeds. Assume for simplicity that the TGA balance is held constant\footnote{This is obviously 
a simplification. In practice, the 
revised cash-balance policy announced by the Treasury in 2015 (see Quarterly Refunding Statement of 5/6/2015, https://www.treasury.gov/press-center/press-releases/Pages/jl10045.aspx), which stipulates holding sufficient
cash to cover a certain number of upcoming daily outflows (subject to a floor), is likely to cause the TGA balance to naturally
drift upward over 
time as the magnitude of average daily outflows increases. Conversely,
at times the debt-ceiling suspension mechanics compels the Treasury to \emph{reduce} the TGA balance dramatically by a
certain date.}  year over 
year.
We therefore must have
\begin{equation}\label{mainrecursion}
N_t = D_t + I_t + M_t
\end{equation}
where $N_t$ is the face notional amount to be issued in year $t$. (Here issuance is assumed to be at par; any issuance discount is assumed to be contained in $I_t$.) More generally, the case where the TGA balance is not constant but rises over time could be addressed, by (for example)
including cash-balance changes into the deficit term $D_t$; as long as 
cash-balance grows at a rate no faster than do deficits, the below analysis is unaffected.

Assume all debt can be issued in one of several tenors $j \in (1, 2, \dots, T)$. Note here all Treasury bills are to be lumped into the $j=1$ bucket; for simplicity the interest on 1-year debt in year $t$ is assumed due in year $t+1$. (As 
well, the inflation component of TIPS and the floating coupons of FRNs are ignored in this construct; where necessary they are lumped in with their nominal counterparts by tenor.)

For our purposes here, an \emph{issuance strategy} is assumed to consist of selecting constant fractions $f_j$ a priori that govern how much of each tenor to issue. That is, in year $t$ Treasury will issue at tenor $j$ the amount
\begin{equation}\label{diffeq}
N_{t,j} = f_j N_t = f_j (D_t+I_t+M_t)
\end{equation}
Obviously the assumption that real-world issuance strategy is described by a set of flow fractions is debatable. In practice, modern Treasury issuance practice has historically been stated and managed in terms of the 
auction sizes (in nominal dollar terms) of its regular notes and bonds. When necessary, changes to those sizes may be considered going forward. The resulting issuance process has, consequently, often been characterized
by punctuated periods during which nominal sizes (and not issuance-fractions) of tenors have remained steady, unless and until changed. This is done
in a deliberate manner in keeping with Treasury's philosophy of "regular and predictable" issuance. Clearly, the (empirical) issuance fractions that result from such a process need not be (and have not been, historically) constant 
or steady. 
However, it will be seen that as an approximation to the actual process over long periods, 
 and as a description of the stance of issuance allocation (as regards its long-term implications), this fractional model will suffice for our purposes.

Because $I_t$ and $M_t$ depend on $N_{s,j}$ for $s \ge t-T$, equation~\ref{mainrecursion} is a $T$-period recursive difference equation for the yearly vectors of new-issue sizes $(N_{t,j})_{j=1}^T$, in which deficits $D_t$ play the role of an exogenous forcing term, one that is presumably
unrelated to debt management policy as such.

One gauge of the cost of this issuance strategy is just its new-issue weighted average maturity (NWAM); if measured in years, this
quantity can be immediately read off of the issuance fractions $f_j$ via the formula
\[
NWAM = \sum (j-\frac{1}{2}) f_j = \sum j f_j - \frac{1}{2}
\]
(The $1/2$ term here adjusts for the fact that  Treasury issuance is typically spread throughout a year.) 

\begin{figure}[htbp]
\includegraphics[scale=.7]{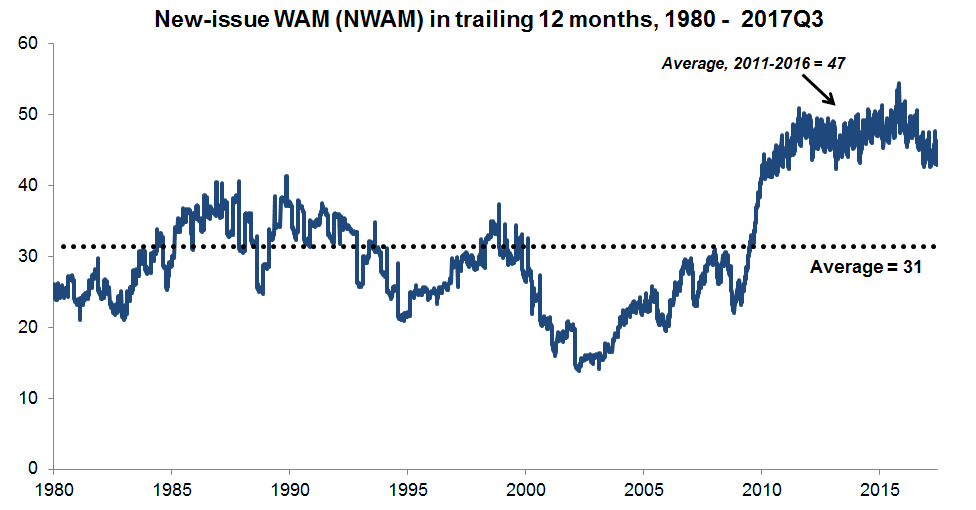}
\caption{NWAM (new-issue weighted-average maturity) of Treasury issuance in trailing 12 months. Since 2010-11 the average maturity of new-issuance has been well above historical averages.  \label{fig_nwam}  }
\end{figure}

In a sense this is the flow equivalent of WAM. This metric itself can provide useful insights into debt management policy. For example, during the period 2010-16 Treasury WAM steadily increased, which has been characterized as a policy of "actively" increasing WAM. But it may
be more descriptive to say merely that Treasury used a very \emph{stable and unchanging new-issuance pattern}, and that pattern
had a longer-tenor bias. This is illustrated by the relatively flat NWAM throughout the period. (See Figure~\ref{fig_nwam}.) The reason WAM increased as a result is that the issuance pattern Treasury set in place in 2010-11, and then left largely unchanged, was long-biased (had a high NWAM of 45-50 months) compared to historical practice and to the (lower) breakeven 
NWAM that would have maintained the length of the outstanding portfolio at its then-extant level.

\section{Steady asymptotic limit}

How is the recursion~\eqref{diffeq} used to derive cost and risk proxies? What we seek is to characterize the \emph{long-term implications}
of using a given strategy, as described by the issuance fractions $f_j$. Conceptually, we would like to think of the portfolio as being in (in some sense) a "steady-state", so that we may ask questions about how current issuance affects or causes the portfolio to drift toward that steady-state. This long-term, steady-state-centered point of view is meant to be the logical extension of a common modeling approach in which a portfolio, under some issuance strategy assumption, is
simulated for a  long period of time (10, 20 years or more) and then the resulting metrics are averaged and examined at the horizon or in 
an abstract future period (cf. \cite{bolder1}).

Granted, it is not obvious that such a steady-state can or does exist. In practice, fiscal deficits and interest rates fluctuate in a correlated
and possibly secular manner; issuance tenors and strategies, meanwhile, can change. The Treasury portfolio historically shows little signs of ever having converged to a steady-state per se. As well, nominal quantities all grow over time, restricting the possible 
quantities that can
even be said to be steady or homogeneities that may exist. A model with the ambition of calculating and describing the "steady-state", long-term consequence of an issuance
strategy evidently must be abstract and hypothetical in nature. Simplifying assumptions must be made.

Here, we make two major assumptions:
\begin{itemize}
\item
Deficits grow geometrically at a constant rate of $g$ per year ($D_t \sim (1+g)^t$).
\item
The yield curve (interest rates $r_j, j=1, \dots, T$) is static and constant yearly.
\end{itemize}

 The reader will object that the above are unrealistically simplified assumptions; they are better conceived of as representing \emph{long-term averages} around which a real market-data path might plausibly fluctuate and to which it might mean-revert. 

The motivation for these assumptions is precisely that these appear to be the conditions necessary for a rolling Treasury portfolio 
to become self-similar
and approach a meaningful "steady-state" as $t \to \infty$. Indeed, under these assumptions we can solve for the long-term asymptotic equilibrium portfolio of equation~\eqref{diffeq} and observe its
properties and metrics. In the below, we highlight two such metrics that naturally represent this asymptotic steady-state portfolio
as regards its cost and risk.

\subsection{Cost proxy} 

Under appropriate and historically plausible conditions on $r$ and $g$, we show in  Appendix~\ref{appendixwac} that the effective interest-cost percentage, or (to abuse terminology slightly, given our approximation of discount bills as
1-period coupon bonds) 
weighted-average coupon (WAC) of the portfolio under the strategy $f$ (defined as $WAC_t = I_t / z_{t-1}$, where $z_{\cdot}$ represents the total debt stock) approaches a constant asymptotic limit
\[
WAC_t \to WAC^* = \sum w_j r_j
\]
where $w_j$ are weights that depend only on $f$ and $g$ (not on $r$):
\begin{equation}\label{wac_weights}
w_j = \begin{cases}
\cfrac{f_j (1-(1+g)^{-j})}{\sum_k f_k (1-(1+g)^{-k})}, & g > 0\\
\cfrac{j f_j}{\sum k f_k}, & g=0
\end{cases} 
\end{equation}
In essence $w_j$ are compounded and
 \emph{growth-adjusted} new-issue fractions. They represent the implication that using issuance strategy $f$ 
has \emph{for interest cost} -- how using allocation $f$, in effect, "samples the yield curve" in the long haul. 

The weights $w_j$ are adjustments of the new-issue fractions $f_j$. Note that for $g>0$ the adjustment (downward) is larger for short tenors than for longer tenors. This reflects the fact that longer-tenor interest rates play a larger role in portfolio WAC when the portfolio is dominated by recent issuance and (consequently) recently-issued longer bonds are present as a significant component of stock. That is precisely the situation when deficit growth $g$ is positive and large:
recent issuance is more prominent.

The quantity $WAC^* = WAC^*(f; g, r)$ serves as a cost indicator of using issuance strategy $f$. Obviously it is not intended as a true calculation of the cost either "right now" or on any particular path; rather, it is a gauge of the expected long-term, structural cost (asymptotically) if the strategy is to be maintained, and if near-term fluctuations in deficits and rates are neglected. But given that Treasury subscribes to a philosophy of "regular and predictable" issuance, and does not attempt to time the market, in our view it is the precisely this 
long-term asymptotic expectation that is the proper orientation when it comes to a high-level cost assessment.

Of course, $WAC^*$ also depends strongly on the particular asymptotic rate assumption $r$. To remove this dependence, we calculate a quantity representing a "WAC-effective" tenor:
\[
t_{WAC} = \sum j w_j
\]
Higher/lower values of $t_{WAC}$ should be associated with higher/lower cost, all else equal (and assuming an upward-sloping and relatively-smooth yield curve $r$).  This makes $t_{WAC}$ -- which is effectively a variant of NWAM, but using compounded and growth-adjusted weights $w$ rather than the issuance fractions $f$ -- a useful rate-assumption-free cost proxy for issuance strategy $f$.

\subsection{Risk proxy}

Opinions and choices 
vary on how to gauge risk in debt-management. One can find in regular use as risk metrics the stochastic variance (i.e. uncertainty, whether conditional or unconditional) of interest-cost, the variability of interest-cost one can expect to observe
unfolding over time, and measures of interest cost such as Cost-at-Risk ($CaR$) and variants that are based on the tails of the distribution of
outcomes (examples for some/all of these can be found in, for
example, \cite{adamo}, \cite{balibek}, \cite{bolder1}, \cite{bolder2}, \cite{brazil}, \cite{das}, \cite{date},
\cite{hahm}, \cite{tbac}). There is also a growing view that the focus of cost quantification ought to be the primary balance (or total deficit) 
rather than  interest cost in isolation (cf. \cite{blommestein}, \cite{missale}, \cite{tbac}). Some sovereigns also place exogenous bounds on simple gross quantities 
such as required auction sizes, gross interest cost, or $Debt/GDP$; these too, where applicable, serve as risk constraints.

Without taking specific a view on the matter, we simply observe that arguably, any/all of 
these are dependent on and correlated with a simple quantity, the \emph{yearly maturing amount} 
(i.e., rollover fraction). After all, the higher a percentage of the portfolio that must be rolled yearly,
\begin{itemize}
\item The more exposed is the portfolio to unanticipated interest-rate movement;
\item The more volatile we can expect the portfolio interest cost to be (locked-in interest costs are, by nature, not volatile over time);
\item The larger tail measures such as Cost-at-Risk should be (again because the portfolio is more exposed to unanticipated spikes in interest rates, to tails in their distribution);
\item The larger the bill issue sizes are (more bills is part and parcel of what creates high rollover).
\end{itemize}
The yearly rollover amount is therefore a simple and useful risk proxy. It is calculable in the asymptotic steady-state setup described above. This quantity $RR_t$, representing the percent of portfolio in year $t$ that matures by $t+1$, is shown in  Appendix~\ref{appendixrr} to approach a constant limit:
\[
RR_t \to RR^* = \sum \tau_j w_j 
\]
where the weights $w_j$ are as defined in~\eqref{wac_weights}, and
\[
\tau_j = \begin{cases}
\cfrac{g}{(1+g)^j - 1}, & g>0 \\
\cfrac{1}{j}, & g=0
\end{cases}
\]
As with $WAC^*$, the measure $RR^* = RR^*(f; g)$ gauges the \emph{long-term, asymptotic, steady-state implication} of using 
issuance strategy $f$, but for risk. Namely, using $f$ indefinitely can be expected to create a portfolio of which the fraction $RR^*$ 
must be rolled yearly. Higher $RR^*$ indicates higher risk; this is our risk proxy metric.

\subsection{Frontier}

\begin{figure}[htbp]
\includegraphics[scale=.7]{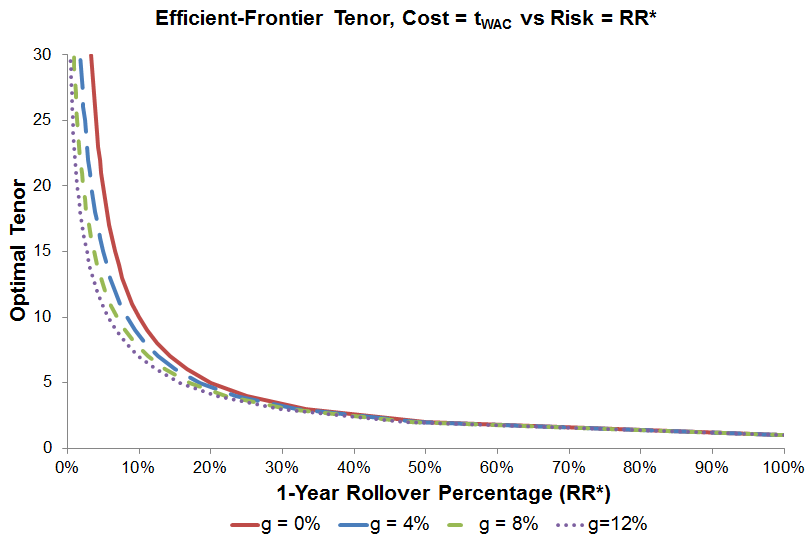}
\caption{"Sweet-spot" tenor vs. rollover ($RR^*$) constraint. Single-issuance strategy at the given tenor (or if not possible, a blend of nearby tenors) dominates
other issuance strategies $f$ with the same $RR^*(f)$. Frontier is shown for $g=0, 4, 8, 12\%$; higher values of $g$ reduce
the risk-mitigation created by longer-tenor issuance, because faster deficit-growth reduces the relative importance of long-past issuance
to the portfolio, making shorter tenors more efficient.  \label{fig_frontier}  }
\end{figure}

A natural question at this point is which strategies comprise the efficient frontier under these cost and risk metrics. In 
Appendix~\ref{appendix2} we show that under certain regularity 
conditions on $r$, when the required risk level is fixed at some given level $\tau_j$ (i.e. the constraint $RR^*  \le R$ is imposed
for $R := \tau_j$), 
the strategy that minimizes the 
cost measure $t_{WAC}$ is issuance concentrated on the single tenor 
\[
j := j^*(R) = 
\begin{cases}
\cfrac{ \log(1+\cfrac{g}{R})}{\log(1+g)}, & g>0 \\
\cfrac{1}{R}, & g = 0
\end{cases}
\]
When $R$ is not of the form $\tau_j$ for integer $j$, but lies in between $\tau_j$ and $\tau_{j+1}$, then the lowest-cost strategy is an 
appropriate blend of issuance at tenors $j$ and $j+1$. So in effect, for any level of risk $R$ these metrics imply there is a "sweet spot" tenor given by the formula $j^*(R)$ around which a concentrated-issuance strategy is the cost-dominant one for risk $= R$. This tenor 
is illustrated in Figure~\ref{fig_frontier} for several choices of $g$.

Note 
in the limit as $g \to 0$ this optimal tenor is simply $j = 1/R$ (see solid curve in Figure~\ref{fig_frontier}). Stated conversely,
when $g=0$ the "risk" of $j$-year issuance using our
metric is just $RR^*(g=0)=1/j$, 
recovering the intuitive notion that a $j$-year issuance strategy creates a portfolio of which roughly $1/j$ is rolled yearly. Indeed, 
\cite{sweden} proposes a risk proxy based on assigning weight $1/j$ to $j-$period borrowing.  By using 
the metric $RR^*$ as risk proxy, we are able to make the appropriate and growth-adjusted generalization of that simple intuition to
arbitrary new-issuance strategies $f$.

\section{Long-term deficit and rate assumptions}

The metrics described above depend on deficit-growth and interest rate assumption parameters. For our purposes it is best to 
calibrate these to longer-term averages that can serve as a plausible description of asymptotic behavior.

 For 
rates we can use, for example, a simple average of history rates (see Table~\ref{table_rates}). Such a yield curve is wider than
the current market environment as of 2017, so it embeds an assumption that some reversion to the mean or normalizing will have taken place in the long-term. Internal calculations use a set of long-term forward rate assumptions produced by OMB as part of its budget process.

\begin{table}[htbp]
  \centering
  \caption{Historically averaged yields 1981-present (derived from H15 curve; source: FRED). }
    \begin{tabular}{ccccccccc}
    Tenor & 1y    & 2y    & 3y    & 5y    & 7y    & 10y   & 20y   & 30y \\
    Yield (\%) & 3.24  & 3.56  & 3.79  & 4.22  & 4.54  & 4.79  & 4.88  & 5.39 \\
    \end{tabular}%
  \label{table_rates}%
\end{table}%

\begin{figure}[htbp]
\includegraphics[scale=.62]{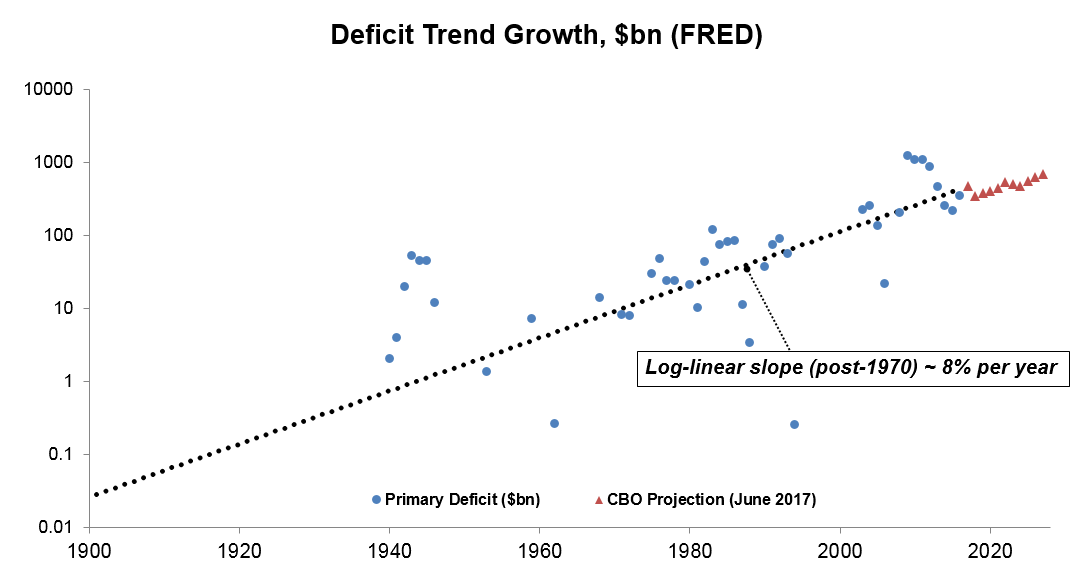}
\caption{Primary deficits (source: FRED) portrayed log-linearly, excluding surplus years. The 
long-term trend (post-1970) is closest to an $8\%$ per year increase. The implied growth of 
future primary deficits and other-means-of-financing forecast by CBO (also shown) as of June 2017 is
consistent with this trend.   \label{fig_deficit8}  }
\end{figure}

For deficits we observe the log-linear trend of deficit growth over several historical decades (post-1970, when primary deficits began to be the
norm) to be roughly $g=8\%$, so that
is our base case choice. See Figure~\ref{fig_deficit8}.

Because the intent is to be parsimonious and represent \emph{average} dynamics rather than to overfit to any particular rate or deficit scenario, the hope is that results are not overly sensitive to these choices. A natural question therefore is how our proxy metrics vary with the rate and deficit assumptions behind them. Obviously $t_{WAC}$ is unaffected by the particular choice of $r$, by design. $WAC^*=\sum w_j r_j$ depends in a linear way on $r$; tilting $r$ wider and steeper would increase $WAC^*$ (and conversely for tighter/flatter), but in a somewhat uniform way. That is, when comparing two strategies in terms of $WAC^*$, simple shifts or rescalings of $r$ (within plausible limits) will 
tend not to materially alter the \emph{relative} cost relationship between them.

The effect of the deficit growth assumption $g=8\%$ may appear more ambiguous. Generally, for a given strategy $f$, 
increasing the assumed $g$ will increase its computed cost ($t_{WAC}$) but decrease its computed risk ($RR^*$). Figure~\ref{fig_frontier}
shows the effect of these shifts for various choices of $g$ on frontier (concentrated-issuance) strategies. As 
with rates, such a shift in $g$ within a plausible range of $4-12\%$ appears unlikely to materially
alter conclusions drawn regarding the \emph{relative} cost/risk relationship between two sufficiently-distinct strategies.

\begin{figure}[htbp]
\includegraphics[scale=.63]{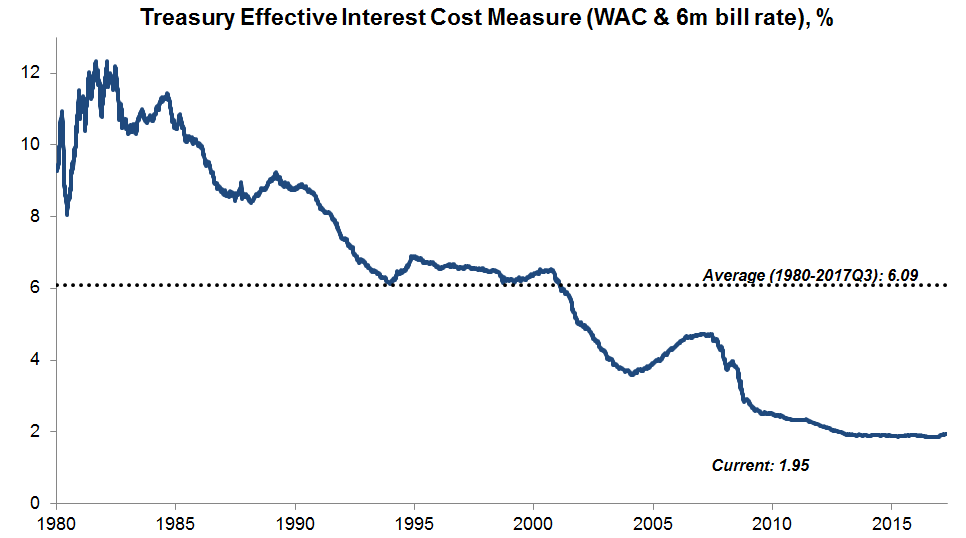}
\caption{Interest-cost measure based on weighted-average coupon (WAC) of outstanding Treasury securities, adjusted by
the $6m$ bill rate $\times$ the bills percentage of the portfolio, from 1980 to 2017Q3. The adjustment 
is necessary because bills contribute a $0\%$ coupon to the naive WAC calculation. 
 \label{fig_wac}  }
\end{figure}

As stated in Appendix~\ref{appendix1}, there is also a technical requirement, for our simple construction to lead to 
the explicit analytical formulas described above, that $g > WAC^*$. (Otherwise the equilibrium $WAC^*$ can 
only be calculated implicitly.) This does not appear to be overly restrictive; as shown in Figure~\ref{fig_wac}, 
using a simple estimate of the \emph{effective} WAC (adjusting the WAC of coupon-bearing securities by the $6m$ bill
yield $\times$ the bill fraction of the portfolio) we see that only in the early 1980s Volcker era did the effective-WAC
exceed $8\%$. (Note, this was also a period when deficit growth was far higher than $8\%$ per year.) The 
average is closer to $6\%$ and the 
recent trend has been under $2\%$. All indications are that deficit-driven rather than interest-driven debt growth is the
asymptotic behavior that is relevant.

It is also worth acknowledging that our calculations are in strict nominal terms and based on the simple mechanical mathematics of debt-rolling. In so doing we knowingly leave considerations
such as the debt-to-GDP ratio and the intertemporal budget constraint, transversality conditions, and/or 
sustainability considerations (see, for example, \cite{ley}) aside. As well, apparently 
the oft-cited correlation
or feedback between issuance, interest rates, and fiscal needs as in the tax-smoothing and Ramsey planning literature (cf. \cite{lucas}, \cite{bohn}, \cite{cochrane}) is omitted here.

\section{Mapping strategy to cost-risk space}

We have described how a set of issuance fractions $f$ can, under simple assumptions about deficits and rates, be mapped into cost-versus-risk space. There are several alternatives described above (depending on the level of exactness or abstraction desired):
\begin{itemize}
\item
$f \to (RR^*, NWAM)$
\item
$f \to (RR^*, WAC^*)$
\item
$f \to (RR^*, t_{WAC})$
\end{itemize}
Summarize this abstractly by stipulating that to each set of fractions $f$ we form the mapping
\[
f \to (R,C)
\]
where $R$ is the risk-proxy metric and $C$ the cost-proxy chosen. (In what follows, unless otherwise stated we use $C=t_{WAC}$.)

To illustrate this mapping, when $f = e_j$ (i.e. concentrated regular issuance only at tenor $j$), this 
mapping becomes
\[
f = e_j \to (\tau_j, j)
\]
The $j=1$ strategy maps to $(R=100\%, C=1)$, low cost but high risk. The $j=30$ strategy maps (with $g = 8\%$) to 
$(R= 0.88\%, C=30)$, high cost but low risk. Other examples of this mapping and its dependence on $g$ for single-tenor strategies $f=e_j$ are 
shown in Table~\ref{tab:single_tenor_mappings}, which also highlights the dependence on $g$, 
especially for longer tenors. (This also helps illustrate why we did not ignore deficit-growth in our construct, or use the naive
rollover metric obtained by simply taking $\tau_j = 1/j$.) 

Note that in this reductive case, the choice of
$r$ plays little meaningful role, as long as $r$ is monotone increasing (as would be the case for any plausible long-term asymptotic
rate assumption). This is because 
using $C=WAC^*$ we just have $WAC^* = r_j$. Since $r_j$ is monotone-increasing in $j$, the two proxies are equivalent
in the sense of sorting single-tenor strategies by relative cost (i.e., $r_j > r_k \iff j > k$ for all $j, k$).

\begin{table}[htbp]
  \centering
  \caption{Examples of the mapping $f \to (R,C)$ where $f$ represents a single-tenor issuance strategy at tenor 
$j=1,2,3,5,7,10,30$. The cost proxy
$C=t_{WAC}$ for this case reduces to $C=j$. (If $C=WAC^*$ is used instead, then we would just have $C=r_j$, which for monotone
increasing rates $r$ would not alter the relative cost picture.) The risk proxy $R$ depends on the deficit-growth assumption $g$: faster
deficit growth implies less relative rollover contribution from long-past issuance of longer tenors.}
    \begin{tabular}{cccc}    
      & \multicolumn{3}{c}{Deficit growth factor (g)} \\
 \underline{   Single-tenor strategy (y) }&  \underline{ 4\%}   &  \underline{ 8\% }  &  \underline{ 12\%} \\
    1     & (100\%,1) & (100\%,1) & (100\%,1) \\
    2     & (49\%,2) & (48.1\%,2) & (47.2\%,2) \\
    3     & (32\%,3) & (30.8\%,3) & (29.6\%,3) \\
    5     & (18.5\%,5) & (17\%,5) & (15.7\%,5) \\
    7     & (12.7\%,7) & (11.2\%,7) & (9.9\%,7) \\
    10    & (8.3\%,10) & (6.9\%,10) & (5.7\%,10) \\
    30    & (1.8\%,30) & (0.9\%,30) & (0.4\%,30) \\
    \end{tabular}%
  \label{tab:single_tenor_mappings}%
\end{table}%

Real-world issuance strategies are of course not usually 
concentrated on single tenors, as for various reasons Treasury can be seen to issue
regularly across the yield curve. But what this method allows is to place any nontrivial issuance pattern $f$ into $(R,C)$ space 
in a way that remains consistent with simple intuition 
about the tradeoffs between cost and risk in the  single-tenor thought experiment. Applications of this principle are demonstrated
in the next section.

\section{Usage and results}

In this section we describe how the proxy metrics described above are used to illustrate and visualize debt issuance strategy. All
projected quantities are hypothetical and for illustrative purposes only.

\subsection{Summarizing historical issuance strategy}

How can these metrics, which by construction relate 
to asymptotic portfolio behavior, be applied to actual Treasury debt issuance? It can be observed historically that Treasury does 
not evidently select and then use indefinitely a single  set of fractions $f$ for its issuance pattern. Instead, the issuance pattern can only be observed empirically, as a consequence of Treasury decisions and events. Nor does the 
asymptotic portfolio ever actually emerge, as fiscal, market and (indeed) debt management strategy may evolve and fluctuate.

But for historical issuance we can still analyze issue fractions a posteriori: aggregate all issuance within a fiscal year and observe the empirical issuance fractions that were used, by setting
\[
f_j = A_j / A
\]
where here 
$A_j$ represents the amount issued in tenor bucket $j$, and $A$ represents total issuance, during the year. (Note that bills
that were issued and matured prior to the end of the fiscal year are not counted in $A_t$, since it is meant to reflect \emph{net} borrowing
required to 
finance $D_t + I_t + M_t$.) Table~\ref{tab:fy2016} shows how this calculation applies to summarizing Treasury issuance during fiscal-year 2016.
\begin{table}[htbp]
  \centering
  \caption{Fiscal-year 2016 Treasury issuance summarized as issuance fractions by tenor bucket. Bills are placed in the 1y bucket, 
TIPS/FRNs with their nominal counterparts. Bills issued/matured within the year are excluded; only net new issuance (live as of $9/30/16$) is included. Figures here exclude debt purchased by Federal Reserve SOMA account as auction add-ons,  being not directly
germane to Treasury issuance strategy.}
    \begin{tabular}{ccccc}
    \toprule
    \multicolumn{5}{|c|}{\textbf{Empirical Issuance Fractions (FY2016)}} \\
    \midrule
    \multicolumn{1}{|c}{\textit{\textbf{Tenor (y)}}} & \textit{\textbf{Security Type}} & \textit{\textbf{Notional (\$bn)}} & \textit{\textbf{Flow}} & \multicolumn{1}{c|}{\textit{\textbf{Issuance Fraction}}} \\
    \midrule
    \multicolumn{1}{|c}{\textbf{1}} & Bills & \$1,647  & \$1,647  & \multicolumn{1}{c|}{\textbf{42.3\%}} \\
    \midrule
    \multicolumn{1}{|c}{\multirow{2}[2]{*}{\textbf{2}}} & Notes/Bonds & \$350  & \multirow{2}[2]{*}{\$520 } & \multicolumn{1}{c|}{\multirow{2}[2]{*}{\textbf{13.4\%}}} \\
    \multicolumn{1}{|c}{} & FRN   & \$170  &       & \multicolumn{1}{c|}{} \\
    \midrule
    \multicolumn{1}{|c}{\textbf{3}} & Notes/Bonds & \$300  & \$300  & \multicolumn{1}{c|}{\textbf{7.7\%}} \\
    \midrule
    \multicolumn{1}{|c}{\multirow{2}[2]{*}{\textbf{5}}} & Notes/Bonds & \$462  & \multirow{2}[2]{*}{\$509 } & \multicolumn{1}{c|}{\multirow{2}[2]{*}{\textbf{13.1\%}}} \\
    \multicolumn{1}{|c}{} & TIPS  & \$47  &       & \multicolumn{1}{c|}{} \\
    \midrule
    \multicolumn{1}{|c}{\textbf{7}} & Notes/Bonds & \$381  & \$381  & \multicolumn{1}{c|}{\textbf{9.8\%}} \\
    \midrule
    \multicolumn{1}{|c}{\multirow{2}[2]{*}{\textbf{10}}} & Notes/Bonds & \$267  & \multirow{2}[2]{*}{\$347 } & \multicolumn{1}{c|}{\multirow{2}[2]{*}{\textbf{8.9\%}}} \\
    \multicolumn{1}{|c}{} & TIPS  & \$80  &       & \multicolumn{1}{c|}{} \\
    \midrule
    \multicolumn{1}{|c}{\multirow{2}[2]{*}{\textbf{30}}} & Notes/Bonds & \$167  & \multirow{2}[2]{*}{\$189 } & \multicolumn{1}{c|}{\multirow{2}[2]{*}{\textbf{4.9\%}}} \\
    \multicolumn{1}{|c}{} & TIPS  & \$22  &       & \multicolumn{1}{c|}{} \\
    \midrule
          & \textit{Total:} & \textit{\$3,892 } &       &  \\
    \end{tabular}%
  \label{tab:fy2016}%
\end{table}%

\begin{figure}[htbp]
\begin{center}
\includegraphics[scale=.45]{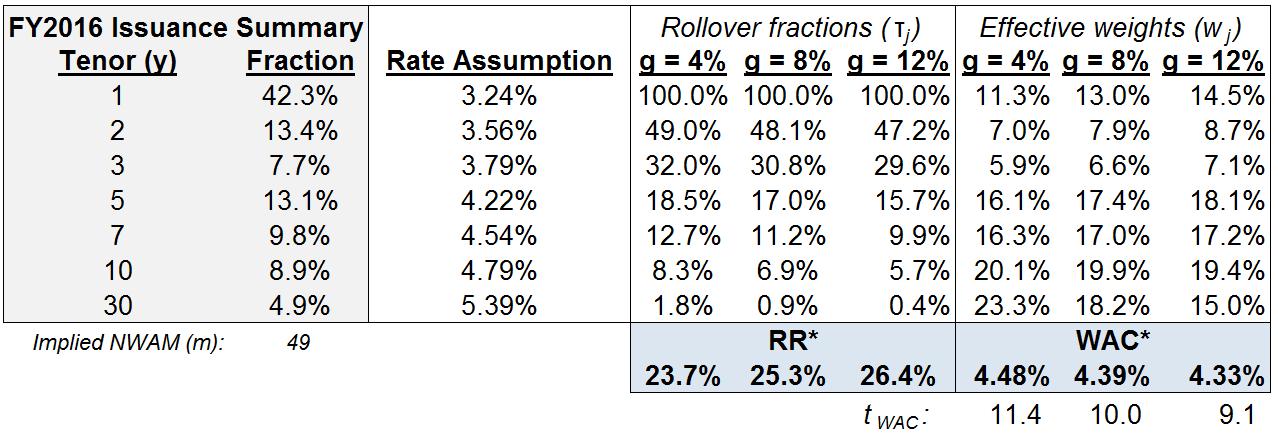}
\caption{ Mapping $f \to (R,C)$ shown for FY2016 issuance summarized as empirical new-issue fractions, and using a long-term
rate assumption derived from post-1980 historical-averages. Risk ($RR^*$) and Cost ($WAC^*$) proxies shown for three 
deficit growth assumptions, $g = 4, 8, 12\%$. Using base-case $g=8\%$, the method summarizes 2016 issuance as 
having selected $(R,C) = ( 25.3\%, 4.39\%)$ in cost-versus-risk space.  Average new-issue maturity was around four years ($NWAM = 49$ months), but $t_{WAC} = 10$ implies that under this issuance pattern it is the $\sim$ 10-year
interest rate behavior/assumption that will be of most importance to longer-term portfolio interest cost when
issuing using this allocation. \label{fy2016RC}}
\end{center}
\end{figure}

We can thus create this mapping in each historical year $t$:
\[
\mathrm{Debt\;issued\;in\;year\;} t \to f^{(t)}  \to (R^{(t)},C^{(t)})
\]
In so doing we are summarizing a year's issuance pattern by mapping it to \emph{the asymptotic properties of the portfolio it would create if continued indefinitely}. The table presented in Figure~\ref{fy2016RC} summarizes 
this calculation when applied to 2016 issuance in fractional form.

\begin{figure}[htbp]
\begin{center}
\includegraphics[scale=.67]{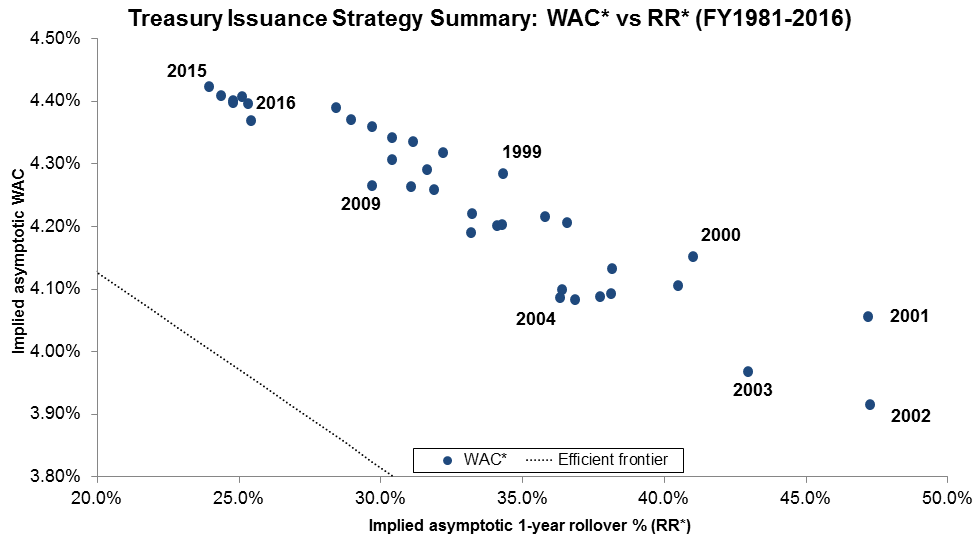}
\caption{Empirical Treasury issuance from FY1981-2016 summarized on a $(RR^*, WAC^*)$ diagram. Assumes 
deficit-growth trend $g=8\%$. Rate assumption derived from historical H15 rates, 1981-2016 (FRED). SOMA 
auction purchases and  Bills issued as part of the Federal  
Reserve's Supplementary Financing Program (SFP) are excluded for this exercise. Omits effect of the large-scale buyback operations in 2000-2002, which were largely focused on 15+ year debt, implying an \emph{effective} issuance policy slightly shorter (higher $RR^*$)
on net than portrayed here.  \label{wacstarsummary}}
\end{center}
\end{figure}

Treasury issuance strategy from fiscal years 1981-2016 is summarized and portrayed in Figure~\ref{wacstarsummary} (using 
$g = 8\%$) with this technique. The approach serves to highlight outlier years: 2015 was an especially long-biased 
issuance pattern; 2001-2002 were short-biased years (in large part due to the discontinuation of 30-year bond issuance). 

We see that current Treasury issuance (FY2016) has pulled back somewhat from its recent long bias. This is presumably a partial 
consequence of the mid-2015 decision\footnote{Quarterly Refunding Statement 
of Acting Assistant Secretary for Financial Markets Seth B. Carpenter, 
5/6/2015 (https://www.treasury.gov/press-center/press-releases/Pages/jl10045.aspx)} to increase bills 
stock, but also of the February 2016 decision\footnote{Quarterly 
Refunding Statement of Acting Assistant Secretary for Financial Markets Seth B. Carpenter, 
2/3/2016 (https://www.treasury.gov/press-center/press-releases/Pages/jl0338.aspx)} to reduce issue sizes of
5y and longer notes/bonds/TIPS.

\begin{figure}[htbp]
\begin{center}
\includegraphics[scale=.67]{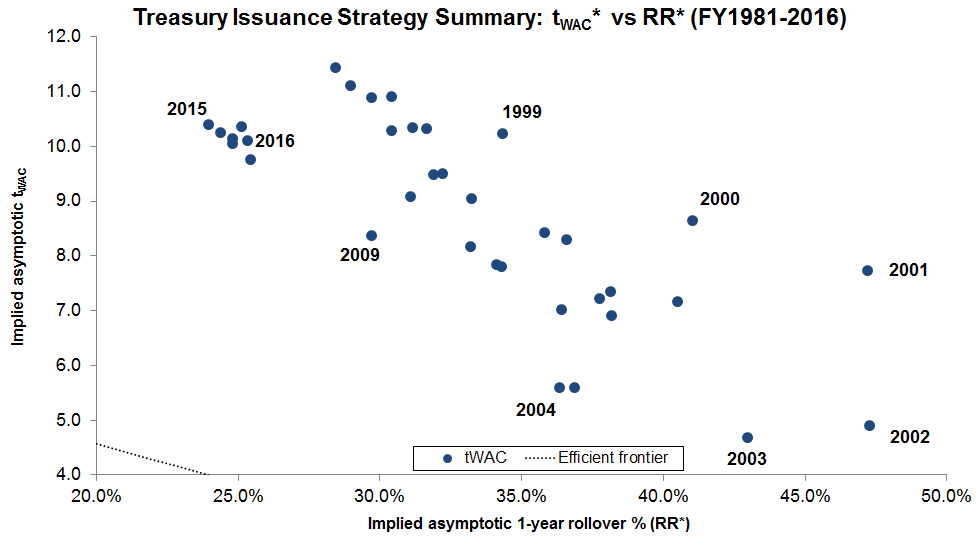}
\caption{Empirical Treasury issuance from FY1981-2016 summarized on a $(RR^*, t_{WAC}^*)$ diagram. Assumes 
deficit-growth trend $g=8\%$. The dependence of the cost metric on rates has been removed by using $t_{WAC}$ as 
cost proxy. Relative relationships between yearly strategies and outliers remain largely unchanged. \label{twacsummary}}
\end{center}
\end{figure}

Figure~\ref{twacsummary} portrays the same diagram but using $t_{WAC}$ rather than $WAC^*$ as cost proxy. 
This transforms and stretches the diagram, but does not appear to materially alter the relative relationships between yearly
strategies, or outliers. This 
should be unsurprising given the fact that $r$ is monotone-increasing (as would be any plausible long-term
asymptotic interest-rate assumption).

\begin{figure}[htbp]
\begin{center}
\includegraphics[scale=.67]{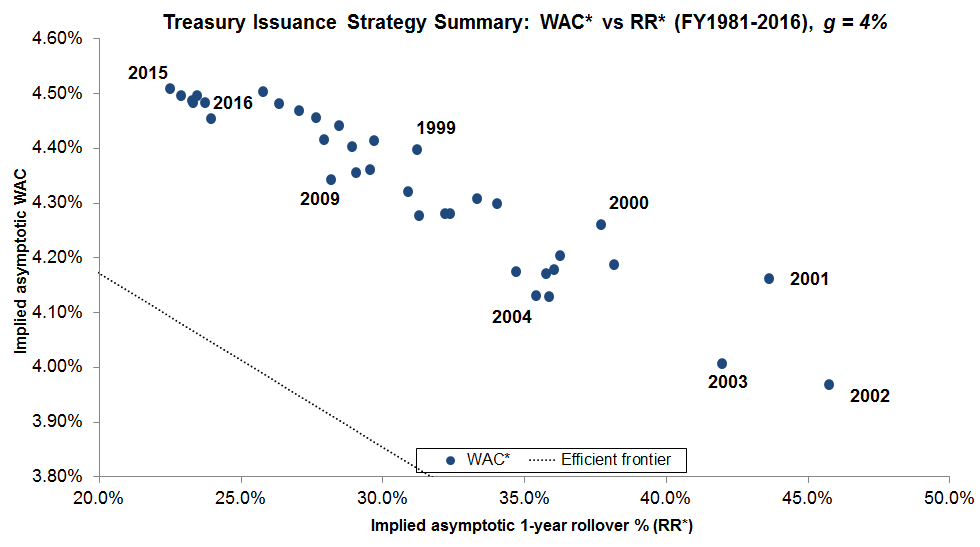}
\caption{Empirical Treasury issuance from FY1981-2016 summarized on a $(RR^*, WAC^*)$ diagram, assuming 
deficit-growth trend $g=4\%$. \label{wacstarsummary_g4}}
\end{center}
\end{figure}

\begin{figure}[htbp]
\begin{center}
\includegraphics[scale=.67]{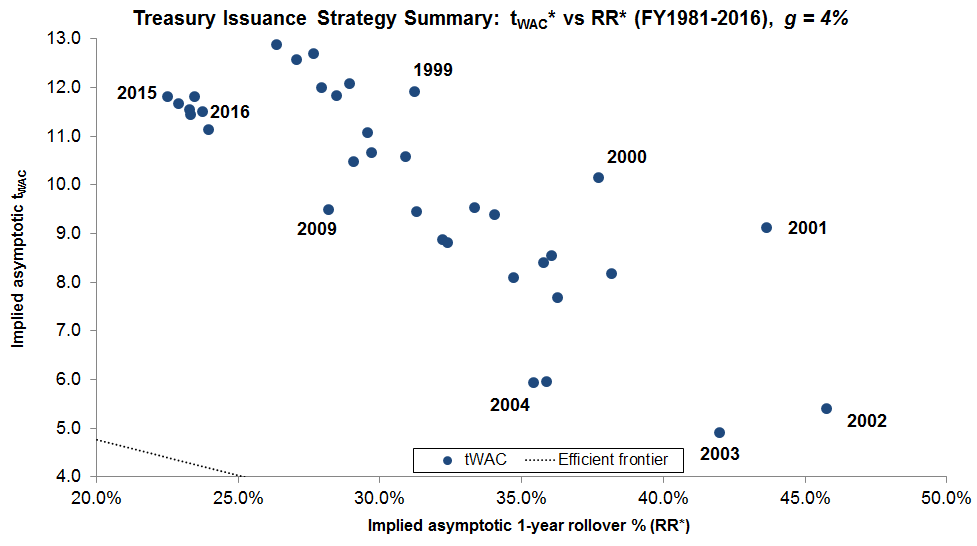}
\caption{Empirical Treasury issuance from FY1981-2016 summarized on a $(RR^*, t_{WAC}^*)$ diagram, assuming 
deficit-growth trend $g=4\%$. \label{twacsummary_g4}}
\end{center}
\end{figure}

Figures~\ref{wacstarsummary_g4}-\ref{twacsummary_g4} replicate these diagrams but using $g=4\%$ rather than $8\%$. We
see that although the quantities change, relative relationships among yearly strategies are largely (albeit not entirely) robust to 
changes in the long-term deficit-growth assumption.

\subsection{Spot issuance strategy}

Rather than the a posteriori empirical analysis of the previous section, we can 
gauge the \emph{current stance} of issuance strategy by using prevailing issuance sizes (rather than rolled-up issuance throughout a year) to infer the new-issue fractions $f$ that they imply. 

\begin{figure}[htbp]
\begin{center}
\includegraphics[scale=.55]{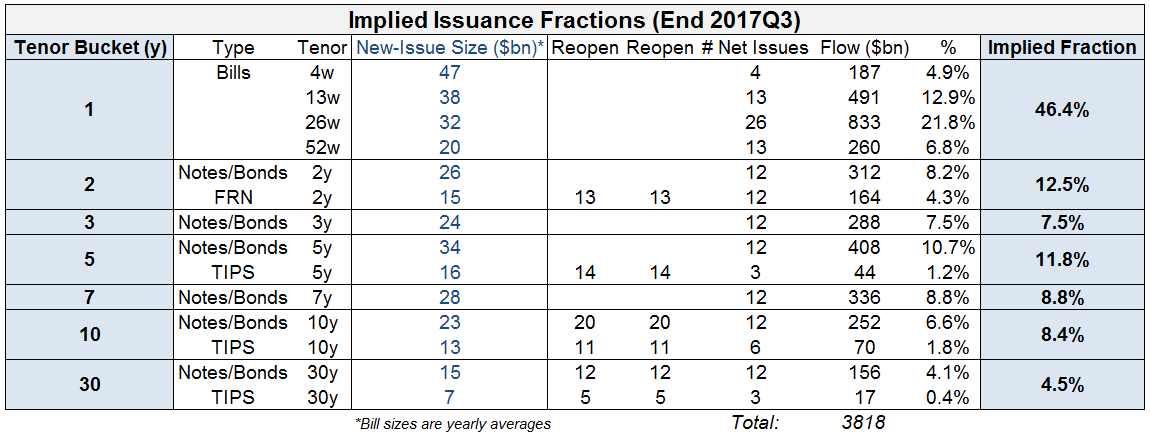}
\caption{Treasury issuance strategy in terms of new-issue fractions 
inferred from new-issue sizes as of end 2017Q3. For bills, the average size over the preceding year is used in order to remove seasonality effects. \label{impliedNI2016}}
\end{center}
\end{figure}

For example, the current size of a new-issue 30-year bond (as of 2017) is $\$15$ billion, a new CUSIP is issued four times per year, and each is reopened twice in amounts of $\$12$ billion. Under such a policy, the implied yearly flow of 30-year issuance is $4 (15 + 12 + 12) = \$156$ billion. We calculate the flow for all other tenors in the same manner, and from those, the implied fractions $f$ these sizes represent. This is illustrated in Figure~\ref{impliedNI2016} for issuance sizes prevailing as of end Q3 2017. (As a technical note: when doing this for bills we use \emph{yearly-averaged} bill sizes rather than current bill sizes, because bill issuance fluctuates throughout the year due to their role as fiscal shock-absorber. Since our goal is to infer the stance of issuance strategy we need to eliminate this exogenous seasonality effect.)

\begin{figure}[htbp]
\begin{center}
\includegraphics[scale=1.05]{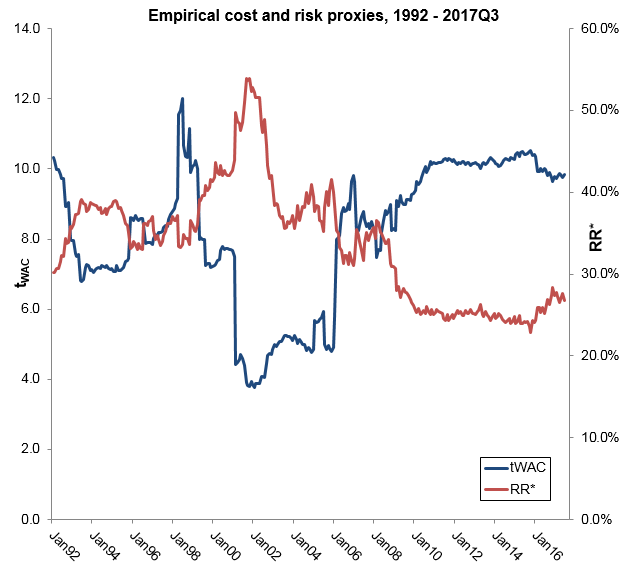}
\caption{Implied monthly cost- and risk-proxies employed for Treasury issuance 
from 1992 - June 2017. Most-recent new issue
sizes are combined with upcoming auction schedules to infer issuance fractions. CMBs, Bills issued as part of the Federal  
Reserve's Supplementary Financing Program (SFP), and effect of 2000-2002 large scale buybacks are excluded for this exercise.
\label{timeGraph}}
\end{center}
\end{figure}

\begin{figure}[htbp]
\begin{center}
\includegraphics[scale=1.05]{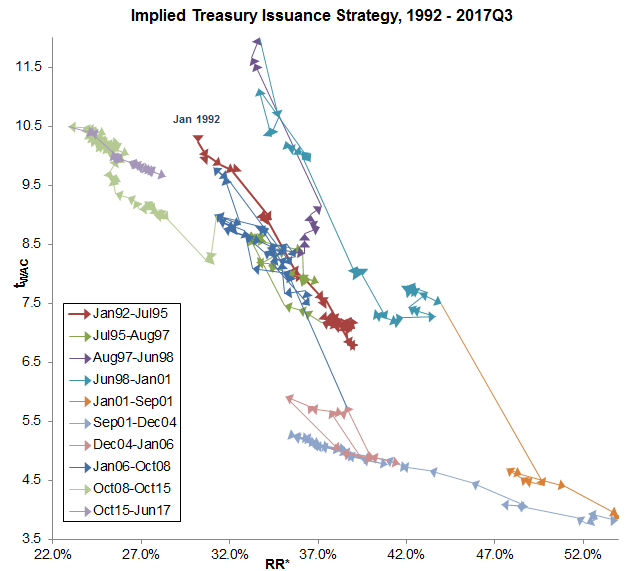}
\caption{Implied monthly path of Treasury issuance 
strategies from 1992 - June 2017 portrayed on an $(R,C)$ diagram. Most-recent new issue
sizes are combined with upcoming auction schedules to infer issuance fractions. CMBs, Bills issued as part of the Federal  
Reserve's Supplementary Financing Program (SFP), and effect of 2000-2002 large scale buybacks are excluded for this exercise.
\label{spotAll}}
\end{center}
\end{figure}

\begin{figure}[htbp]
\begin{center}
\includegraphics[scale=0.51]{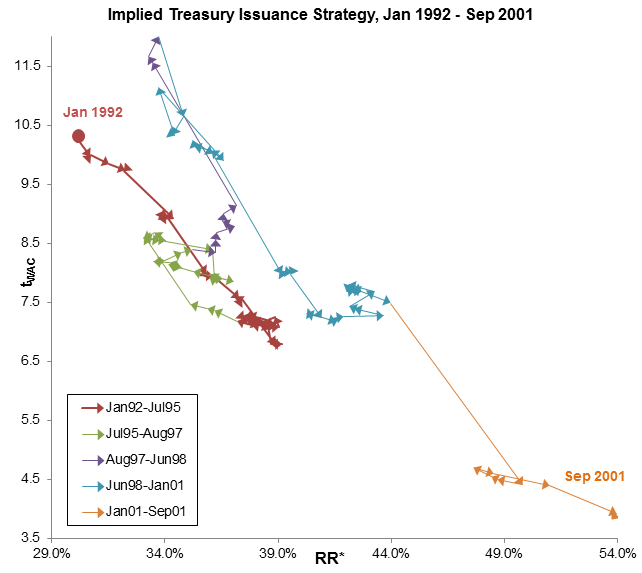}
\includegraphics[scale=0.51]{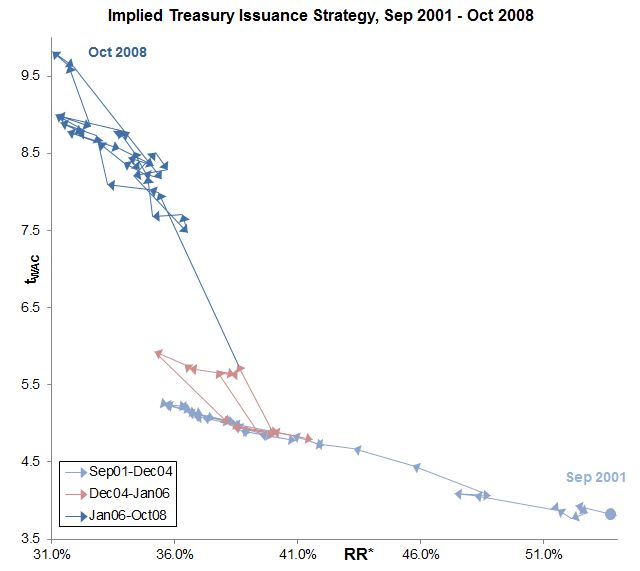}
\includegraphics[scale=0.51]{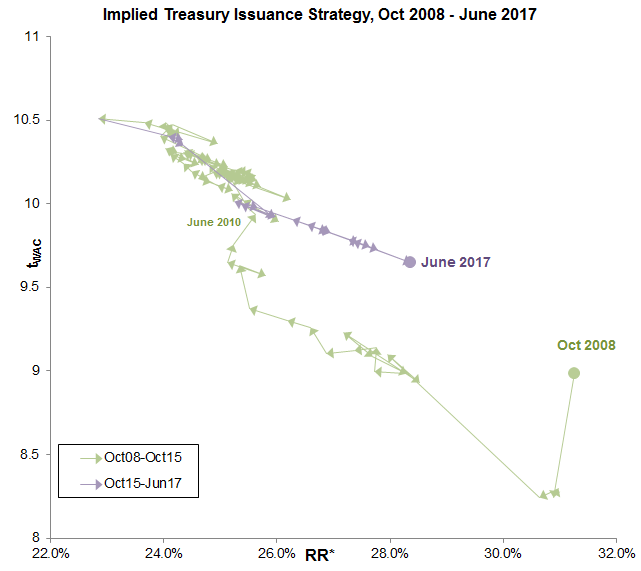}
\caption{ Treasury Issuance strategy 1992 - 2017Q3 broken out into three periods: pre-9/11/2001, pre-financial crisis,
and post crisis. 2010-2016 implied strategy shows as relatively constant (green cluster) in the third graph. \label{spot13}}
\end{center}
\end{figure}

Spot strategies are calculated monthly in this manner from 1992-present (see Figure~\ref{timeGraph}). 
The 
joint evolution is portrayed in Figure~\ref{spotAll} as a trajectory on an $(R,C)$ diagram.  They 
are also broken out into individual periods in Figure~\ref{spot13}. Several apparently 
distinct periods of Treasury issuance policy can be tracked on these illustrations:
\begin{itemize}
\item Reduced term borrowing from falling deficits in 1990s
\item Elimination of 30y bond in 2001 - 2006
\item Shift in term borrowing post 9/11/2001
\item Lengthening of issuance allocation post-financial crisis
\end{itemize}
What is perhaps most salient here, in the final graph of Figure~\ref{spot13}, is 
the clustering of policy points (green, upper left) 
that is seen from June 2010 - October 2015. 
This is the post-financial-crisis period of actively extending WAM, i.e., of employing a relatively static
new issuance flow strategy with a long bias. The 
graph also shows that policy has begun to shorten modestly since, 
with the early 2016 reductions in all 5+ year issuance sizes.

Applying $(R,C)$ metrics to spot-issuance patterns allows one to illustrate and visualize the path and direction of Treasury issuance in a way that lines up with how debt management is most directly announced 
and affected: by considering and/or
making specific changes to issue sizes, by introducing or removing tenors, by changing auction schedules -- by altering Treasury \emph{flow}. Key historical decisions about Treasury issuance become plainly visible as movement in $(R,C)$ space. Conversely, periods such as 2011-16 in which (we assert) issuance strategy was effectively unchanged emerge as periods of little or no such movement. Note that in such periods of static issuance-strategy, WAM may be changing rapidly, which again
highlights the distorted picture that can be painted by WAM.

\subsection{Visualizing forward scenarios}

Another benefit of mapping issuance to our representative $(R,C)$ metrics is that one can visualize the direction and effect of potential forward strategy options in a simple, intuitive way that escapes the need to examine dozens or hundreds of individual, projected future
issue sizes. 

For example, suppose financing needs are due to increase, and Treasury considers one of three strategy options in response:
\begin{itemize}
\item Increase only bill sizes incrementally as needed; leave all other sizes unchanged
\item A shorter-tenor strategy preferentially favoring increases to bills, 2-3y notes, and the 2y floating rate note
\item Keep bill, TIPS and FRN sizes constant, and increase the coupon stack pro rata as needed
\end{itemize}

\begin{figure}[htbp]
\begin{center}
\includegraphics[scale=0.5]{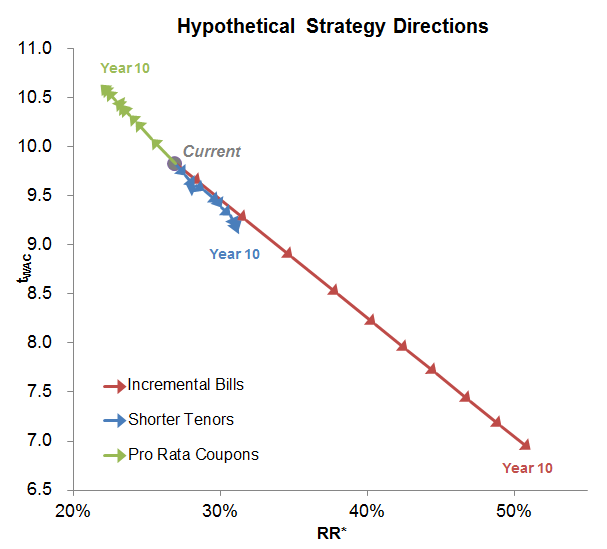}
\caption{Yearly spot-$(R,C)$ diagrams of three 
hypothetical strategy approaches to financing incremental issuance over a 10-year forward period. \label{hypothetical3}}
\end{center}
\end{figure}

In Figure~\ref{hypothetical3} each of these three strategy approaches have been simulated for ten years under hypothetical rate and deficit assumptions, and the issue-size-implied allocations
plotted as trajectories on the $(R,C)$ diagram. The different cost and risk implications of each of the strategy options become
evident.

Portraying strategies in this way can help the policy maker understand the implications of and relationships between the cost-risk tradeoffs involved in strategy alternatives.

\subsection{Constrained scenarios}

The idealized efficient frontier mentioned above in $(R,C)$ space will typically consist of single-tenor issuance at some optimal 
tenor. Obviously this is not realistic, practical, or even desirable: Treasury has a long-established policy of issuing and maintaining
many liquid benchmarks across the yield curve. There are other factors that can motivate against the literal interpretation of the concentrated issuance as optimal: the model omits feedback between issuances and rates, for
example. For all intents, a concentrated 
issuance strategy $f=e_j$ would be considered inadmissible as a strategy option. This raises the 
point that strategies must belong to the \emph{admissible set} of strategies in order to be legitimately considered as potential allocation adjustments. 

\begin{figure}[htbp]
\begin{center}
\includegraphics[scale=0.5]{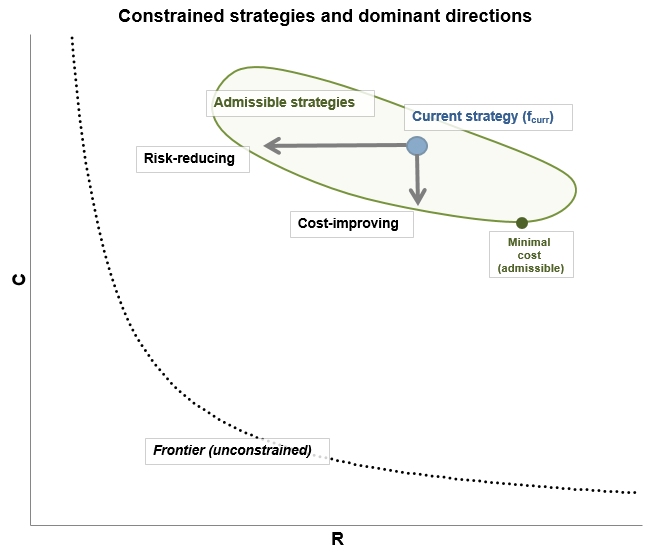}
\caption{Schematic illustration of issuance strategies constrained to a "policy window" (admissible set) of allowed 
strategies, mapped to an $(R,C)$ diagram. Cost- and risk-dominant strategy directions from the current issuance
pattern $f_{curr}$, as well as the optimal strategy within the admissible set (generally a simplex, for simple
linear constraints), are easily calculated. \label{pwcartoon}}
\end{center}
\end{figure}

Defining admissible strategies is ultimately the province of the debt manager; the process and considerations behind it are beyond the scope of this piece. But the effect of constraints -- the debt manager's "policy window" of admissible strategies --
on the analysis above, modifies the framework  only slightly. A common approach may be to define lower and upper
bounds $L$ and $U$ for the new-issuance allocations (or equivalently, upper bounds on changes from the current allocation
$f_{curr}$); this defines the admissible set to be strategies such that 
$L \le f \le U$.  Each such strategy has a point on the $(R,C)$ diagram, and it is straightforward to identify dominant directions
that reduce cost or risk (see Appendix~\ref{appendix3}). This idea is depicted in Figure~\ref{pwcartoon}.

These dominant and constrained-optimal strategy directions could be used to inform
policy decisions as to how to adjust borrowing patterns in the face of anticipated changes in financing requirements. For example,
if deficits are anticipated to rise (creating a "funding gap" if current issue-size flow were maintained), borrowing at 
which tenor(s) should be increased in order to close the gap? Different choices of tenor(s) imply different new-issue allocations
$f$, which can be evaluated in terms of how they alter the cost and risk proxy metrics described above.

\section{Summary}

We introduce simple cost and risk proxy metrics that can be attached to empirical Treasury issuance, to "spot" issuance patterns, and/or to forward issuance scenarios over time. These metrics are based on mapping issuance fractions to their long-term, asymptotic portfolio implications for cost ($C$) and risk ($R$) under 
mechanical debt-rolling
dynamics, and given simple but necessary assumptions about asymptotic deficit growth and interest rates.

Mapping strategies to the resulting $(R,C)$ diagram enables one to harness the intuitions of standard portfolio theory, including the efficient frontier, which in this simple construct can be calculated analytically as an optimal single-tenor issuance strategy corresponding to a given risk tolerance. 

Comparing historical or future issuance to the frontier and to past issuance enables policy makers to understand the direction and implication of the issuance decisions and implied tradeoffs they make in a way that is more dynamic and flow-centered than the traditional portfolio metric (WAM). As such these metrics are a valuable tool in discussing and understanding Treasury debt management.

Although the recursion model used to derive these metrics is clearly reductive, there is scope to expand on the work by, for example, endowing it with rate and deficit paths that are stochastic, mean-reverting, and correlated. Ultimately the steady $WAC^*$  and rollover-fraction $RR^*$ that we derive under our simple steady recursion model may be understood to represent -- under appropriate assumptions -- a special case of
the stochastic equilibrium portfolio of a more complex and market-realistic model. Another potential area of refinement is to model the dynamics of inflation-linked (TIPS) and floating-rate (FRNs) debt rather than binning them with nominal counterparts. 

While work is ongoing, we hasten to add that in bridging the gap between debt management modeling and practice  model complexity and (often illusory) macroeconomic completeness ought not be impediments to intuitive metrics that can aid consideration and communication of debt management strategy -- and, which are aligned with the considerations of debt management as it is
implemented.

\appendix
\section{Appendix}

\subsection{Equilibrium portfolio distribution} \label{appendix1}

Start with equation~\eqref{diffeq} for new-issue amounts issued yearly, written as a vector equation for 
$n_t = (N_{t,j})_j$:
\[
n_t = f( D_t + I_t + M_t)
\]
Define the \emph{stock} in year $t$:
\[
z_t = z_{t-1} + N_t - M_t = z_{t-1} + D_t + I_t 
\]
Using our asymptotic assumption about deficits we can write 
\begin{equation} \label{asymD}
D_t = D_0 \gamma^{t-1} 
\end{equation}
where $\gamma := 1+g$. 

We seek the steady average interest cost at equilibrium, defined as 
\begin{equation} \label{asymWAC}
WAC^* = \lim_t I_t / z_{t-1}
\end{equation}
This means asymptotically we should have
\[
z_t = \beta z_{t-1} + D_0 \gamma^{t-1} 
\]
where $\beta := 1+ WAC^*$. This can be solved:
\[
z_t = z_0 \beta^t + \frac{D_0}{\gamma-\beta}(\gamma^t -\beta^t)
\]
Now \emph{assume} that $\gamma > \beta$ (i.e. that $g>WAC^*$). Under this assumption (which we acknowledge to depart from 
what is often assumed in the literature, cf. \cite{barro}, but which appears to be the empirically relevant case), as 
$t \to \infty$ this becomes
\begin{equation}\label{asymz}
z_t \sim \frac{D_0}{\gamma - \beta}\gamma^t,
\end{equation}
i.e. $z_t \sim O(\gamma^t)$.

Define $q_{j,t}$ to be the stock of \emph{current}
tenor $j$ at time $t$. Of course, $z_t = \sum_j q_{j,t}$. Collect these quantities into a vector:
$Q_t := (q_{j,t})_j$. Stock of current-tenor $j$ comes from two sources: rolldown of current-tenor stock $j+1$ from the previous
year, and new-issuance at tenor $j$ in the current year. This implies a recurrence for $Q_t$:
\[
Q_t  = S Q_{t-1} + f(D_t + I_t + M_t)
\]
where $S$ is a shift operator ($S_{ij} = \delta_{j,i+1}$). Notice that the maturing amount $M_t$ is simply
\[
M_t = q_{1,t-1} = e_1^T Q_{t-1}
\]
Therefore
\begin{equation}\label{Qdiffeq}
Q_t = (S+fe_1^T)Q_{t-1} + f(D_t + I_t) := RQ_{t-1} + f(D_t+I_t)
\end{equation}
where $R:=S+fe_1^T$.

We are interested in the evolution of portfolio fractions, defined as 
\[
\theta_t := \frac{1}{z_t}Q_t
\]
That is, $\theta_{j,t}$ represents the percentage of the portfolio at time $t$ with current-tenor $j$. From equation~\eqref{Qdiffeq}
we find
\[
\theta_t = \frac{1}{z_t} Q_t = \frac{z_{t-1}}{z_t} (\frac{1}{z_{t-1}} Q_{t-1}) + \frac{1}{z_t} f(D_t + I_t)
=\frac{z_{t-1}}{z_t} R\theta_{t-1} + \frac{1}{z_t} f (D_t +I_t) := A+B+C
\]
and treat the terms $A$, $B$, and $C$ separately by using~\eqref{asymz}, \eqref{asymD} and \eqref{asymWAC} to write
\[
A \to \frac{1}{\gamma} R\theta_{t-1}
\]
\[
B \to \frac{1}{\gamma}(\gamma-\beta)
\]
\[
C \to \frac{1}{\gamma}(\beta-1)
\]
The equilibrium portfolio distribution $\theta_t \to \theta^*$ must therefore satisfy
\[
\theta^* = \frac{1}{\gamma} R\theta^* + (1 - \frac{1}{\gamma} ) f
\]
or
\[
\theta^* = (\gamma-1) (\gamma I - R)^{-1} f
\]
Letting $y = \theta^*/(\gamma-1)$ we must have $(\gamma I - R) y = (\gamma I - S)y - y_1 f = f$, or 
$y = (1+y_1)(\gamma I - S)^{-1} f$. Equivalently
\[
\theta^* \propto (\gamma I - S)^{-1} f := T_{\gamma} f
\]
where the constant of proportionality is pinned down by the condition that $\sum \theta^*_j =1 $. So we may write
\[
\theta^* = \frac{1}{ || T_{\gamma} f ||_1} T_{\gamma} f
\]
and it is easy to show that
$T_{\gamma}$ is an upper-triangular matrix with
\[
(T_{\gamma})_{ij} = \frac{1}{\gamma^{j-i+1}}, j\ge i
\]
The preceding holds in the limiting case $\gamma \to 1$ as well.

\subsection{Asymptotic WAC}\label{appendixwac}

Assume the interest-rate applicable to tenor-$j$ issuance in year $t$ is simply $r_j$. This is paid out in the following years 
$t+1, t+2, \dots t+j$ as an amount $r_j N_{j,s}$. Therefore the net interest owed in year $t$ from all prior-year issuance of
tenors $j=1, \dots, M$ is
\[
I_t = \sum_{j=1}^M \sum_{s=t-j}^{t-1} r_j N_{j,s} = \sum_{j=1}^M r_j f_j ( \sum_{s=t-j}^{t-1} N_s)
\]
In the limit we seek, $I_t = WAC^* z_{t-1} = (\beta-1) z_{t-1} $, or
\begin{equation}\label{trywac}
I_t \sim \frac{1}{\gamma} (\beta-1) z_t
\end{equation}
using the asymptotic behavior~\eqref{asymz} of $z_t$. To derive the appropriate $WAC^*$ rewrite the 
above as
\begin{equation} \label{ib}
I_t = ( \sum_{j=1}^{M} r_j f_j b_{j,t} ) z_t
\end{equation}
where
\begin{equation}\label{bjt}
b_{j,t} := \frac{1}{z_t} \sum_{s=t-j}^{t-1} N_s
\end{equation}
Asymptotically, expression~\eqref{diffeq} for the sequence $N_t$ can be written 
\[
N_t = D_0 \gamma^{t-1} + (\beta - 1) \left[ z_0 \beta^t + \frac{D_0}{\gamma-\beta}(\gamma^t -\beta^t) \right]
+ \sum_{j=1}^M f_j N_{t-j}
\]
where we have symbolically inserted asymptotic expressions~\eqref{asymD} and \eqref{asymWAC} for $D_t$ and $I_t$ 
respectively, and made use of the fact that the maturing amount relates to prior issuance by
$M_t = \sum_{j=1}^M f_j N_{t-j}$.

This is a nonhomogeneous difference equation for $N_t$ of the form
\[
N_t = \sum_{j=1}^M f_j N_{t-j} + a_t
\]
where asymptotically, and using our assumption of $\gamma > \beta$,
\begin{equation}\label{as}
a_t \sim D_0 \gamma^{t-1} \frac{\gamma-1}{\gamma - \beta}.
\end{equation}
It may be homogenized by defining $U_t := N_t - C a_t$, using
\[
C := \frac{1}{  \sum_j f_j (1- \gamma^{-j})}
\]
Then $U_t = \sum_{j=1}^M f_j U_{t-j}$; in particular, the sequence $U_t$ is bounded (because $\sum f_j =1 $). As a result,
$U_t / z_t \to 0$ as $t\to \infty$ and can be neglected. We are left with $0 = N_t - C a_t$ or $N_t \sim C a_t$; 
returning to~\eqref{bjt} we find asymptotically,
\[
b_{j,t} \sim \frac{1}{z_t} C \sum_{s=t-j}^{t-1} a_s \sim C \frac{1-\gamma^{-j}}{\gamma},
\]
where we have used expression~\ref{as} for $a_s$ and~\ref{asymz} for $z_t$. Expression~\eqref{ib} becomes
\[
I_t = \frac{1}{\gamma} C \left( \sum_{j=1}^M r_j f_j (1-\gamma^{-j}) \right) z_t
\]
Comparing with~\eqref{trywac}, we identify
\[
WAC^*= \beta - 1 = C \sum_{j=1}^M r_j f_j (1-\gamma^{-j})  = 
\frac{  \sum_{j=1}^M r_j f_j (1-\gamma^{-j}) }{  \sum_j f_j (1- \gamma^{-j})} = \sum w_j r_j,
\]
with 
\begin{equation}\label{wwac}
w_j = \frac{ f_j (1-\gamma^{-j})}{ \sum_k  f_k (1-\gamma^{-k}) }.
\end{equation}
Note $\sum w_j = 1$ and so the $w_j$ have an interpretation as portfolio weights; $WAC^*$ is a weighted-sum of 
the assumed interest-rate curve $r$.

\subsection{Asymptotic yearly rollover}\label{appendixrr}

The yearly rollover fraction we seek is the limit of
\[
RR_t := q_{1,t}  / z_t = \theta_{t,1} 
\]
From the preceding this approaches the steady quantity $RR^* := \theta^*_1 = (T_{\gamma} f)_1 / || T_{\gamma} f||_1$.

Rewrite expression~\eqref{wwac} for the coupon-effective weights $w_j$ as $w_j = c_j f_j$ with 
\[
c_j \propto 1-\gamma^{-j}
\]
and $\sum w_j = 1$. Since all $c_j > 0$ we can just as well
write this as $f_j \propto d_j w_j$ for $d_j := 1/ (1-\gamma^{-j})$, as long as we again ensure that $\sum f_j = 1$. 
This gives 
\[
f = D_{\gamma} w / || D_{\gamma} w||_1
\]
where $D_{\gamma}$ is a diagonal matrix with $j^{th}$ entry $d_j$. The equilibrium portfolio written this way becomes
\[
\theta^* = T_{\gamma} D_{\gamma} w / || T_{\gamma} D_{\gamma} w||_1
\]
and so we have 
\[
RR^* = \theta^*_1 = 
\frac{e_1^T  T_{\gamma} D_{\gamma} w}{\mathrm{1}^T  T_{\gamma} D_{\gamma} w} 
\]
For the denominator, it is easy to show that $\mathrm{1}^T  T_{\gamma} D_{\gamma}  = \mathrm{1}^T$; and 
since $\mathrm{1}^T w = 1$ for
valid weights $w$, it need not be written. Similarly, simple algebra shows 
\[
\tau^T := e_1^T T_{\gamma}D_{\gamma} = (1, (\gamma-1)/(\gamma^2-1), 
(\gamma-1)/(\gamma^3-1), \dots, 
(\gamma-1)/(\gamma^M-1) )
\]
(Note that as $\gamma \to 1$ this approaches 
the more intuitive limit, $\tau \to (1, 1/2, \dots, 1/M)$.) We therefore have,
\[
RR^* = \tau^T w
\] 
where $\tau_j= (\gamma-1) / (\gamma^j - 1)$. This is the expression (with $\gamma-1 = g$) used in the text.

\subsection{Frontier, optimal tenor, and no-barbell condition} \label{appendix2}

Given issuance strategy $f$, deficit-growth assumption $\gamma$, and rate assumption $r$, we 
now have expressions for the asymptotic portfolio cost
\[
C(w) = r^T w
\]
and risk
\[
RR(w) = \tau^T w
\]
using our chosen proxies, where $w = w(f)$ are accumulated portfolio weights ($\sum w_j = 1$, $w_j \ge 0$) that relate to new-issue allocations $f (\sum_j f_j = 1, f_j \ge 0)$ via
\[
w_j \propto c_j f_j\]
and
\[
c_j = 
\begin{cases}
1 - \gamma^{-j}, & \gamma>1 \\
j, & \gamma=1
\end{cases}
\]

The frontier is characterized by minimizing $C(w)$ subject to a constraint on $RR(w) = R \in (0,1]$. Here we derive conditions on the interest
rate curve $r$ sufficient to ensure a unique optimal strategy, and show that it is characterized by concentrated issuance on a single tenor or combination of adjacent tenors.

\myhed{Candidate optimal strategy}

First, since the function defining $\tau$ above is monotone and onto $(0,1]$, we can write $R = \tau_j$ for some $j$, i.e. $j = \tau^{-1}(R)$. Trivially, when $j  \in \mathbb{Z}$, the single-tenor
strategy $w^*:=e_j$ has risk $RR(w^*)=\tau_j = R$ and cost $C(w^*)=r_j$. When $j \notin \mathbb{Z}$,  set 
$w^* := (1-\theta) e_i + \theta e_{i+1}$ where $i := \floor{j}$. The choice of $\theta = (\tau_j - \tau_{i})/(\tau_{i+1}-\tau_{i})$ ensures $RR(w^*) = \tau_j = R$. The cost of this strategy is 
$C(w^*) = (1-\theta) r_i + \theta r_{i+1}$.

The goal is to show that no other strategy $w$ with the same risk ($RR(w) = \tau_j$) has $C(w) \le C(w^*)$ as given above.

\myhed{Solving the Lagrangian}

The Lagrangian for this problem is
\[
L(w,\lambda,\eta,\alpha)=r^T w + \lambda (\tau^T w - \tau_j) + \eta (\mathrm{1}^T w - 1) + \alpha^T w
\]
where $\lambda, \eta$ and $\alpha \in \mathbb{R}^M$ are Lagrange multipliers associated with the constraints $\tau^T w = \tau_j$, $\sum_l w_l = 1$, and $w_l \ge 0$ respectively. The latter constraints are not binding, and so we may have 
$\alpha_l=0$, in which case $w_l \neq 0$ 
(and vice versa).

The first-order conditions are the  $M$ equations
\[
0 = \nabla_w L = r + \lambda \tau + \eta \mathrm{1} + \alpha
\]
for $M+2$ unknowns $\lambda, \eta$ and $\alpha$. For any tenor $l$ that is part of issuance ($w_l \neq 0$) we have $\alpha_l = 0$, which means 
\[
r_l = -\lambda \tau_l - \eta
\]

Assume $w \neq w^*$ and $RR(w) =\tau_j$. For $j \in \mathbb{Z}$, since $w \neq e_j$, and if $w=e_k$ for $k\neq j$ it would have suboptimal cost or risk, we can assume there are at least two tenors $m<n$ with $w_m,w_n \neq 0$. We know further that we can choose them such that $m<j<n$ because otherwise, either $C(w) > C(e_j)$ or $R(w) > R(e_j)$ due
to the opposite monotonicities of $r, \tau$. By similar reasoning, when $j \notin \mathbb{Z}$, we can choose $m\le i$ and $n \ge i+1$ -- not both binding, since $w \neq w^*$ -- so that again, $w_m,w_n \neq 0$ and $m<j<n$.

Letting $l = m,n$ in the first-order conditions above pins down the values of $\lambda,\eta$
\[
\lambda = -\frac{ r_m - r_n}{\tau_m-\tau_n},
\eta = \frac{ r_m \tau_n - r_n \tau_m}{\tau_m-\tau_n}
\]
At an extremum the cost function then must be
\[
C(w) = \sum_{w_l \neq 0} r_l w_l = \sum_{w_l \neq 0} (-\lambda \tau_l - \eta) w_l = -\lambda \tau_j -  \eta,
\]
where we have used the constraints $\tau^T w = R = \tau_j$, and $\sum w_l=1$. Substituting for $\lambda, \eta$ we see
\begin{align}
C(w) &= (\frac{ r_m - r_n}{\tau_m-\tau_n}) \tau_j -   \frac{ r_m \tau_n - r_n \tau_m}{\tau_m-\tau_n} \nonumber \\
 &= \frac{1}{\tau_m-\tau_n} \left( ( \tau_j  - \tau_n) r_m + (\tau_m -   \tau_j ) r_n   \right)  \nonumber \\ 
 & = C(w^*) +  \frac{1}{\tau_m-\tau_n} \left( (\tau_j - \tau_n) (r_m - C(w^*) ) + (\tau_m -  \tau_j ) (r_n - C(w^*) )      \label{eq1}     \right)
\end{align}

\myhed{Case 1: $j \in \mathbb{Z}$}

When $j \in \mathbb{Z}$, then $C(w^*) = r_j$ and so expression~\eqref{eq1} is just
\[
C(w) = r_j +  \frac{1}{\tau_m-\tau_n} \left( (\tau_j - \tau_n) (r_m - r_j) + (\tau_m - \tau_j ) (r_n - r_j)\right)
\]
Since $t_m - t_n>0$, the above expression suffices to show $C(w) > r_j = C(w^*)$ if we ensure that 
\begin{equation}
(\tau_j - \tau_n) (r_m - r_j) + (\tau_m - \tau_j ) (r_n - r_j) > 0  \label{crossprod}
\end{equation}
for any possible choices of $m<j<n$. Rearranging, this becomes (emphasizing the functional forms of $r$ and $\tau$)
\[
\frac{ r(j)- r(m) }{ \tau(j) - \tau(m)} > \frac{ r(n) - r(j) }{ \tau(n) - \tau(j) }
\]
Below we will derive sufficient conditions on the convexity of $r$ to ensure that this is the case for all possible $m<j<n$, which will prove that $C(w)>C(w^*)$.

\myhed{Convexity condition}

Because $\tau>0$ is bijective, we can employ a change of variables by writing $s = \tau(t)$ for any $t>0$ (keeping in mind that increasing $t$ corresponds to decreasing $s$). Define $s_j = \tau(j)$ and similar for $m,n$. Consider the expression 
\[
\frac{ r(j)- r(m) }{ \tau(j) - \tau(m)} = \frac{ r( \tau^{-1}( s_j )) - r(\tau^{-1} ( s_m )) }{ s_j - s_m} := D[m,j] 
\]
Because $r, \tau^{-1}$ are smooth, this is the first derivative of the function $\phi(s) := r( \tau^{-1} (s) )$ at some intermediate point $\xi_{m,j} \in (s_j,s_m)$:
\[
D[m,j] = \phi'( \xi_{m,j})
\]
Similarly, the right side of the condition becomes
\[
D[j,n] = \phi'( \xi_{j,n})
\]
for some $\xi_{j,n} \in (s_n,s_j)$. In particular, $\xi_{j,n} < s_j < \xi_{m,j}$. To show $D[m,j] > D[j,n]$ as desired, it therefore suffices to show that $\phi'(s)$ decreases as $s$ decreases: that $-\frac{d}{ds} \phi'(s) < 0$, or $\phi''(s) > 0$.

But $\phi(s) = r(\tau^{-1}(s))$, so 
\[
\phi'(s) = r'( \tau^{-1}(s) ) / \tau'( \tau^{-1}(s) )
\]
and
\[
\phi''(s) = \left[ \tau'( \tau^{-1}(s) )  r''( \tau^{-1}(s) )  / \tau'( \tau^{-1}(s) ) -   r'( \tau^{-1}(s) )  \tau''( \tau^{-1}(s) ) /  \tau'( \tau^{-1}(s) ) \right] / ( \tau'( \tau^{-1}(s) ) )^2
\]
The denominator is everywhere positive, so it suffices to ensure that the numerator of this expression (now recalling $s=\tau(t)$) is too:
\[
\tau'(t) r''(t)/\tau'(t) - r'(t) \tau''(t) / \tau'(t) = r''(t) - r'(t) \tau''(t) / \tau'(t) > 0
\]
Recalling that $r'(t) > 0$, we easily rearrange this to the sufficient \textbf{convexity condition} on $r$:
\[
r'' / r' > \tau'' / \tau'
\]
Recalling that $\tau''>0$ and $\tau'<0$, note that this amounts to a negative lower bound on the convexity of $r$ (i.e. an upper bound on its concavity), as desired. This confirms the intuition that extreme concavity in $r$ could lead to a barbell (highly 
separated two-tenor
issuance) being optimal, but as long as $r$ is not too concave, the frontier (for $R=\tau_j, j \in \mathbb{Z}$) consists only of single-tenor issuance.

\myhed{Case 2: $j \notin \mathbb{Z}$}

If $j \notin \mathbb{Z}$ the numerator of expression~\eqref{eq1} instead involves the adjacent rate values $r_i$ and $r_{i+1}$. We can split it into
\begin{gather*}
(1-\theta) \left[ 
(\tau_j - \tau_n)(r_m - r_i) + (\tau_m - \tau_j ) (r_n - r_i) \right] + \\
\theta  \left[ 
(\tau_j - \tau_n)(r_m - r_{i+1}) + (\tau_m - \tau_j ) (r_n - r_{i+1}) \right] := A+B
\end{gather*}
where $\theta \in (0,1)$. Writing $\tau_j = \tau_i + (\tau_j-\tau_i)$ and some rearrangement converts $A$ into
\[
A = (1-\theta) \left( [(\tau_i-\tau_n)(r_m-r_i) + (\tau_m-\tau_i)(r_n-r_i)] + (\tau_i-\tau_j)(r_n-r_m) \right) := (1-\theta)[ A_0+A_1]
\]
Similarly, we can use $\tau_j = \tau_{i+1} + (\tau_j-\tau_{i+1})$ to convert $B$ into
\[
B = \theta \left( [ (\tau_{i+1}-\tau_n)(r_m - r_{i+1}) + (\tau_m - \tau_{i+1})(r_n - r_{i+1})] - (\tau_j - \tau_{i+1})(r_n-r_m) \right) := \theta[B_0+B_1]
\]
The previously derived convexity condition $r''/r' > \tau'' / \tau'$ ensures $A_0,B_0 \ge 0$ (with equality only if $m=i$ or $n=i+1$, which cannot both be true since $w \neq w^*$), as 
they are identical to expression~\eqref{crossprod} with $i$ and $i+1$, respectively, playing the role of $j$. It remains to
examine the residual $(1-\theta)A_1 + \theta B_1$. But
\[
(1-\theta)A_1 + \theta B_1 = (r_n-r_m) [ (1-\theta) (\tau_i-\tau_j) - \theta (\tau_j - \tau_{i+1}) ]=0
\]
recalling the definition of $\theta = (\tau_j - \tau_i )/(\tau_{i+1}-\tau_i)$. We therefore again have $C(w) > C(w^*)$, showing that the concentrated issuance 
$w^* = (1-\theta) e_i + \theta e_{i+1}$, for $i = \floor{j} = \floor{ \tau^{-1}(R)}$ is the unique optimal strategy.

\myhed{Expressions for convexity condition}

When $\gamma=1$, then $\tau(t)=1/t$ and so the above convexity condition requires
\[
r''(t) / r'(t) > -2/t
\]
The boundary of this condition would be an interest-rate curve of the form $r(t) = A - B/t$.

When $\gamma>1$, then $\tau(t) = (\gamma-1)/(\gamma^t-1)$. The condition reduces to
\[
r''(t)/ r'(t) > \log\gamma \left( \frac{ -\gamma^t - 1 }{ \gamma^t - 1} \right)
\]
It can be verified that when $t > 1$ this approaches, from below, the preceding bound $-2/t$ as $\gamma \to 1$. Hence $r'' / r' > -2/t$ is a  sufficient, albeit more restrictive, condition for any $\gamma$.

While of course this condition may be violated (to say nothing of outright curve inversion) by 
the spot curve on any given day, we have found that for plausible 
long-term, asymptotic rate assumptions of interest, such as those drawn from averaged yield curves or from NSS modeled 
curves, this 
convexity condition is met. Of course, if it is violated 
--  if the 
long-term yield-curve assumption $r$ has regions of sufficient negative-convexity (e.g. has "corners", or 
overall is steep at the short end 
but with a rapid switch to flat at the long end) -- it is easy to see that the preceding concentrated-issuance characterization of the optimal frontier need not hold. However, it can be argued that curves with such kinks or corners are not natural candidates for
a long-term, steady yield-curve.

\myhed{Counterexample}

For example, consider the following upward-sloping rate curve $r$:
\[
r(t) = 
\begin{cases}
t & (t \le 2) \\
2+\epsilon(t-2) & (t > 2)
\end{cases}
\]
(or a smooth approximation of this) for some small $\epsilon>0$, assume for simplicity $\gamma=1$, and impose a
risk constraint of $RR(w) \le 1/2$. The concentrated single-tenor strategy 
$w=e_{2}$ satisfies this constraint as it has $RR(w)=1/2$, and its 
cost is $C(w) = 2$.

Meanwhile, consider a barbelled strategy formed by combining tenors $1$ and $30$, $w_{b}:=\phi e_1+(1-\phi) e_{30}$.
If $\phi :=14/29$ this will have the same rollover ratio, $RR(w_{b})=1/2$. Meanwhile its cost is
\[
C(w_{b})=\phi + (1-\phi) [ 2 + 28 \epsilon] = 1 + \frac{15}{29}(1+28\epsilon).
\]
If $\epsilon < 1/30$ then $C(w_b)<2 = C(w)$, showing that concentrated issuance is not optimal. Here the convexity condition on 
$r(t)$ is violated 
due to the corner around $t=2$. Stated differently, the extreme flatness of the 2s30s curve in this 
idiosyncratic case afforded a benefit to issuing longer.

\subsection{Constrained optimization} \label{appendix3}

Suppose constraints on issuance take the form of lower and upper bounds $L$ and $U$ placed on $f$; that is,
\[
0 \le L_j \le f_j \le U_j \le 1
\]
 for all $j$. Notice we also (trivially) have $U_j = 0$ for tenors $j$ that are not part of Treasury issuance. 

This constrains the problem described in the previous section~\ref{appendix2} but its structure and solution approach
otherwise remains 
unchanged:
\begin{itemize}
\item Find weights $w$ that minimize $WAC^*(w) = w^T r$,
\item Subject to  constraints
\begin{gather*}
\sum w_j = \mathrm{1}^T w = 1 \\
\sum \tau_j w_j = \tau^T w \le R \\
L \le f(w) = D_{\gamma} w / || D_{\gamma} w ||_1 \le U
\end{gather*}
\end{itemize}

The preceding is straightforward to cast as a 
linear programming problem. Optimal weights $w^*$ are easily obtained numerically, from 
which they are then converted into
the optimal new-issue strategy $f^*$ via
\[
f^* = \frac{ D_{\gamma} w^* }{||D_{\gamma} w^*||_1}
\]
Comparison of $f^*$ with the current implied strategy $f$ allows one to identify the cost-dominant direction of improvement.
Repeating this procedure for multiple choices of $R$ one can trace the (constrained) efficient frontier. 


\end{document}